\documentclass{elsart3} 

\usepackage{graphicx}

\begin{document}


\begin{frontmatter}
\title{The first version Buffered Large Analog Bandwidth (BLAB1) ASIC for 
high luminosity collider and extensive radio neutrino detectors}
\author[UH]{L.~Ruckman},  
\author[UH]{G.~Varner\corauthref{varner}}, and
\corauth[varner]{Corresponding author. Tel.: +001 808-956-2987.}
\ead{varner@phys.hawaii.edu}
\author[UH]{A.~Wong}

\address[UH]{Department of Physics and Astronomy, University of Hawaii,
  2505 Correa Road, Honolulu HI 96822, USA}


\begin{abstract}
Future detectors for high luminosity particle identification and ultra
high energy neutrino observation would benefit from a digitizer
capable of recording sensor signals with high analog bandwidth and
large record depth, in a cost-effective, compact and low-power way.\\
A first version of the Buffered Large Analog Bandwidth (BLAB1) ASIC
has been designed based upon the lessons learned from the development
of the Large Analog Bandwidth Recorder and Digitizer with Ordered
Readout (LABRADOR) ASIC.  While this LABRADOR ASIC has been very
successful and forms the basis of a generation of new, large-scale
radio neutrino detectors, its limited sampling depth is a major
drawback.  A prototype has been designed and fabricated with 64k deep
sampling at multi-GSa/s operation.  We present test results and
directions for future evolution of this sampling technique.\\
\end{abstract}


\end{frontmatter}


\section{Introduction}
Observation of the early universe through neutrino messengers of the
highest possible energies requires a detector of enormous instrumented
volume~\cite{SalSA}.  At the same time, lepton flavor identification
of such radio detection events represents a completely unique tool for
the study of cosmological evolution of the universe.
Particle interactions at extreme energies provide a probe capable of
illuminating the completely unknown acceleration mechanisms of the
highest energy cosmic ray events~\cite{SandS}.

Particle identification is also crucial to the physics program of a
next generation ``Super'' B Factory.  Such an accelerator will produce
B mesons in sufficiently copious quantities to permit detailed
scrutiny of standard model predictions in the flavor
sector~\cite{superb}.  Any new theories for physics beyond the
standard model must leave fingerprints that can be detected via flavor
transformation of particles in the final state.  Therefore, particle
identification is essential and the detector and readout electronics
must survive the very high signal occupancies
expected~\cite{supersvd}.

We present results from a deep-sampling ASIC that meets these
requirements, based upon extension of the successful LABRADOR
ASIC~\cite{LAB3}.


\section{Architectural Details}

\begin{figure}[htb]
\includegraphics[width=7.62cm]{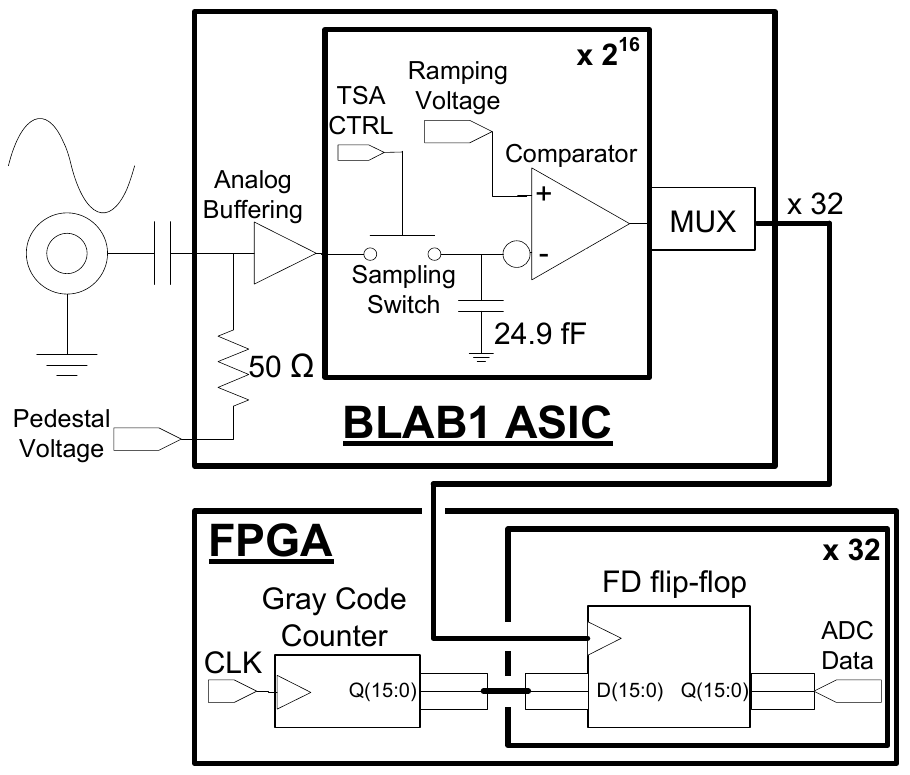}
\caption{A block diagram of the BLAB1 readout, where for compactness
the comparators are located inside the BLAB1 device, while the
high-speed time encoding is done inside a companion FPGA.}
\label{BLOCK}
\end{figure}

The BLAB1 ASIC is a single channel, multi-GSa/s waveform sampler with
a record depth of $2^{16}$ analog storage samples.  The BLAB1 analog
input is AC coupled with an external capacitor and 50 $\Omega$
terminated with an on-chip terminator, as should be expected for a
high-performance RF device.  After the on-chip terminator, an analog
buffer tree fans out copies of the signal to the matrix of 128 rows of
512 samples composing the 64k array.  Each of the rows may be
independently addressed to initiate a storage cycle.  Within each
Switched Capacitor Array (SCA) storage cell is a capacitor and a
comparator. A block diagram of the BLAB1 readout is shown in
Fig.~\ref{BLOCK}.

\begin{figure}[ht]
\vspace*{0mm}
\includegraphics[width=7.2cm]{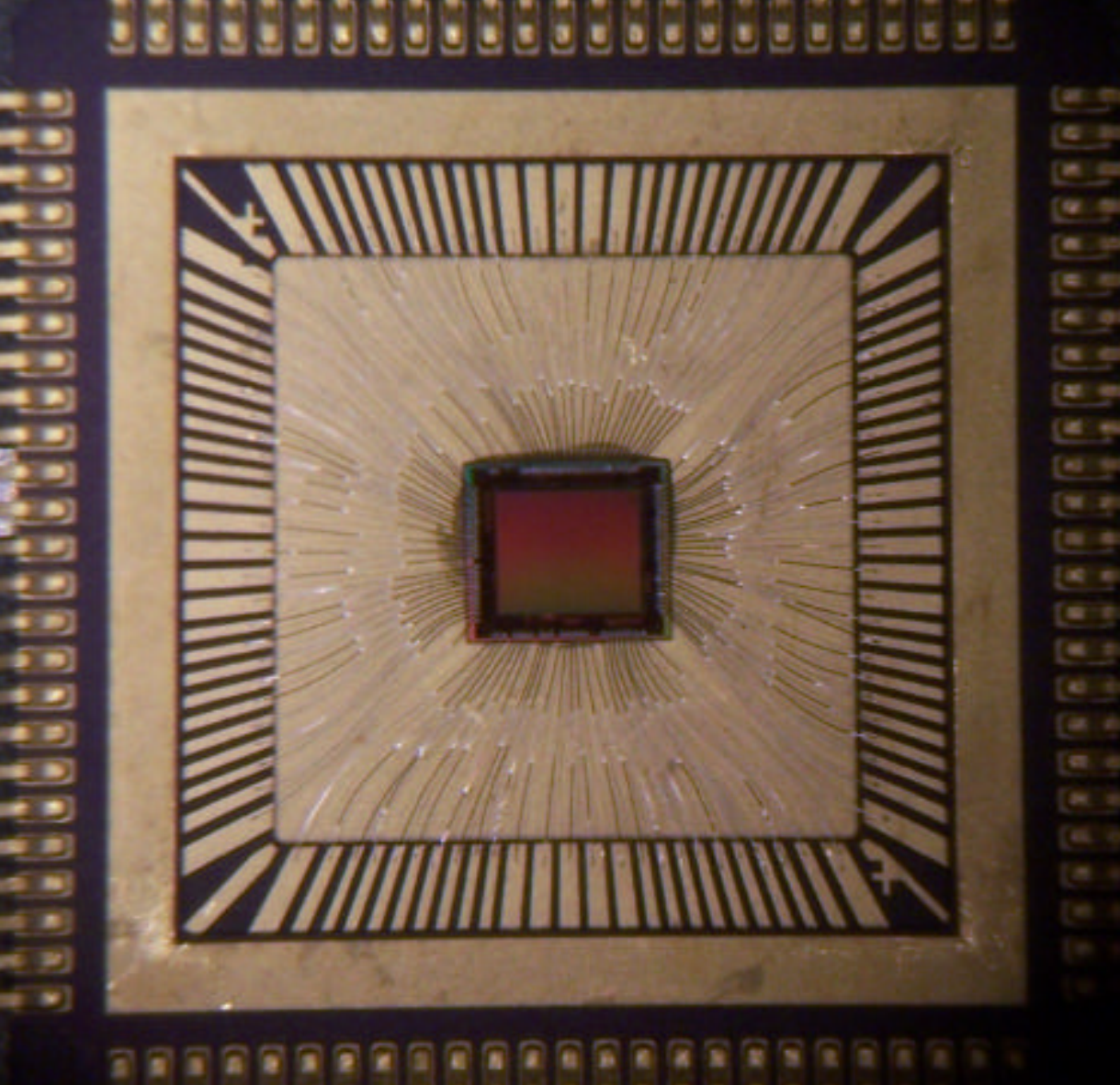}
\vspace*{0mm}
\caption{A photograph of the wire-bonded BLAB1 ASIC.  The die is 3 mm
by 2.8 mm and was fabricated in the TSMC 0.25$\mu $m process.}
\label{bond}
\end{figure}

When an analog switch is pulsed closed, the instantaneous input signal
is stored on a 14 fF capacitor.  The charge is then held until
either overwritten or discharged due to leakage current.  Each
sampling capacitor is connected to the negative input of a comparator.
The positive input of each comparator is connected to a common voltage
ramp.  A wire-bonded BLAB1 die photograph is shown in Fig.~\ref{bond},
with this storage array contained within about 5.25 square mm of
the die shown.

\begin{figure}[ht]
\vspace*{0mm}
\includegraphics[width=8.1cm]{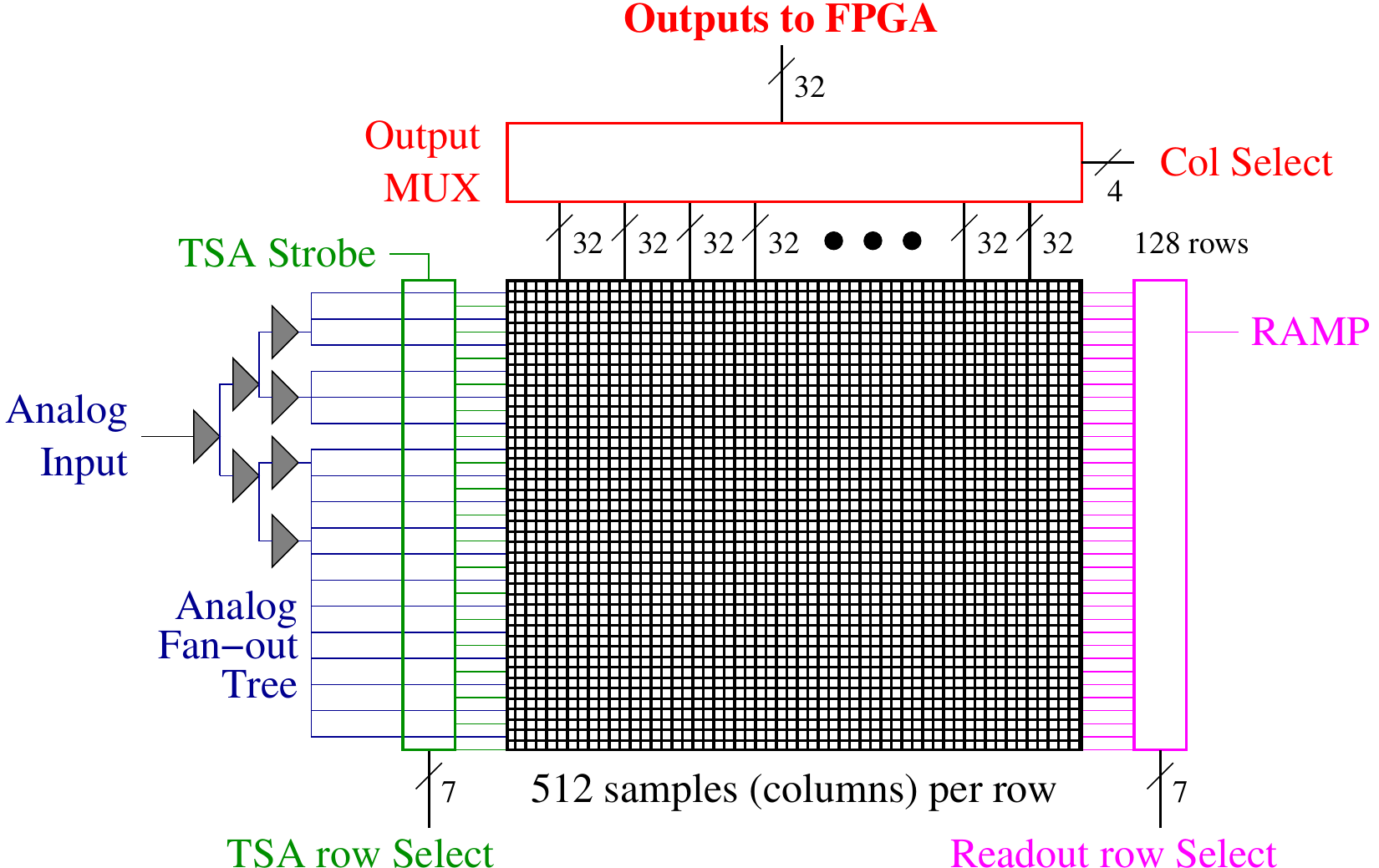}
\vspace*{0mm}
\caption{Schematic of the BLAB1 sampling array.}
\label{array_scheme}
\end{figure}

\indent Conversion of these stored samples is via a Wilkinson ADC
method, where the stored voltage is converted into a transition time
of the in-cell comparator due to an applied voltage ramp.  This ramp
is generated with a current mirror and can be adjusted both by varying
the ramping current, as well as an external capacitor.  The typical
ramping current range is 10-100$\mu $A and the ramp capacitor size is
a few hundred $pF$.  Encoding is performed by measuring the time
interval between the ramp start and the comparator output transition.
In a simple form of time-to-digital conversion, this interval is
measured by counting the number of high-speed clock cycles taken.  In
the predecessor ASIC \cite{LAB3}, the Gray code counter was
implemented on-chip, whereas in BLAB1 it is implemented inside a
companion programmable logic device, in this case a Field Programmable
Gate Array (FPGA).  When the voltage ramp is started, a Gray code
counter in the FPGA is enabled coincident to a high speed clock
(500MHz) and the comparator output is used to latch the counter value.
By knowing the ramping voltage slope and the high speed clock
frequency, the latched counter value can be converted into voltage. A
group of 32 comparators are selected, as illustrated in
Fig.~\ref{array_scheme}, and are read out during each ramping cycle.

\begin{figure}[ht]
\vspace*{0mm}
\includegraphics[width=8.1cm]{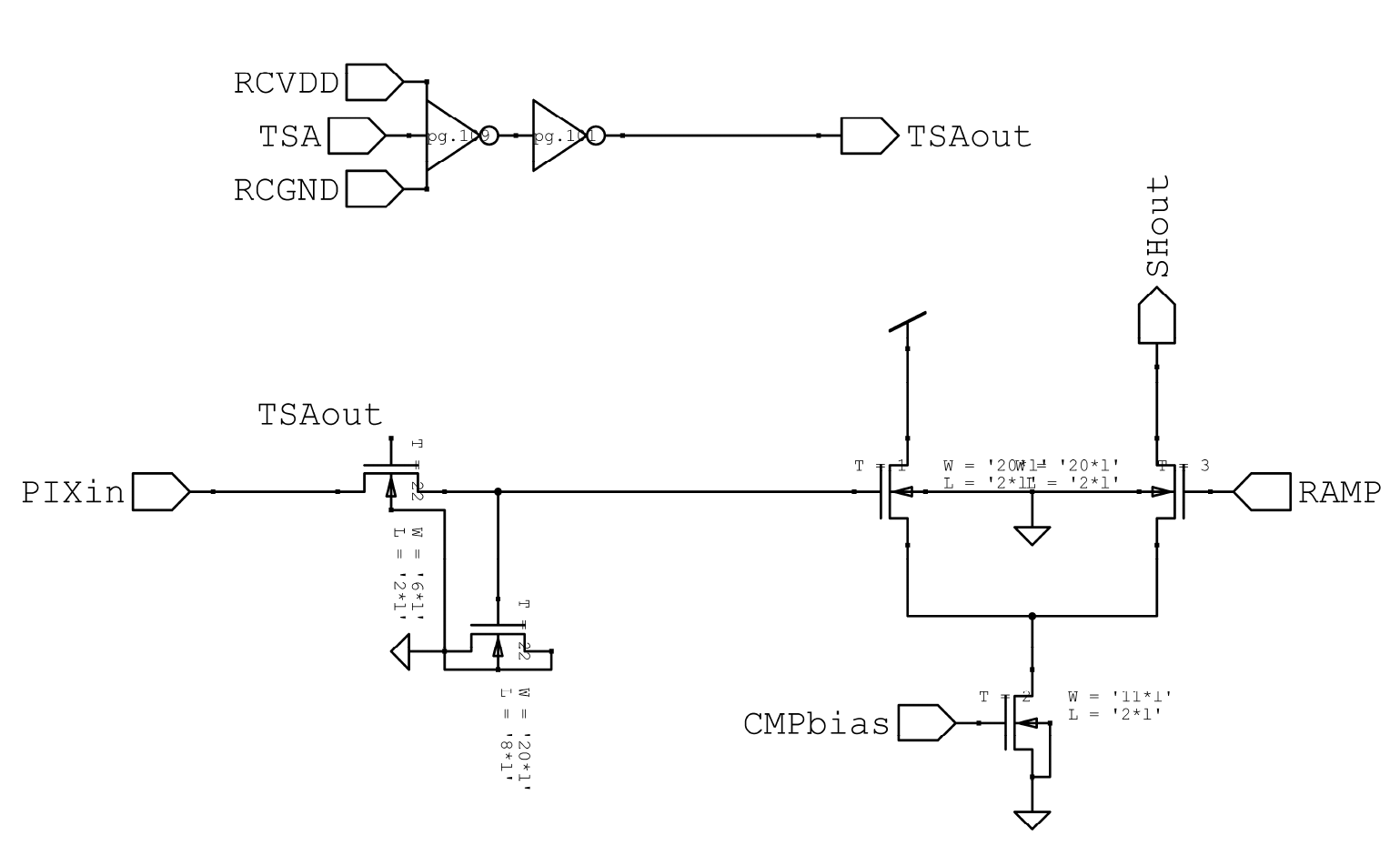}
\vspace*{0mm}
\caption{Schematic of the BLAB1 basic storage cell.}
\label{scheme}
\end{figure}

\begin{figure}[ht]
\vspace*{0mm}
\centerline{\includegraphics[height=9.4cm]{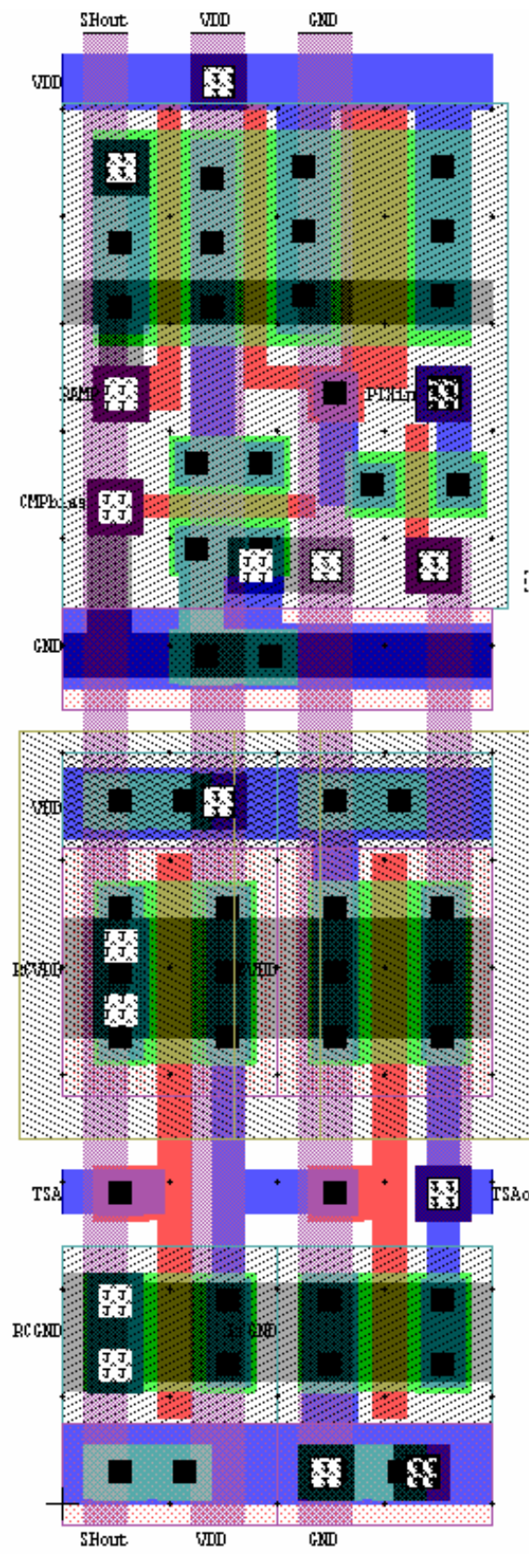}}
\vspace*{0mm}
\caption{Layout of a single SCA storage cell, where the units
displayed are in units of $\lambda$, which is $0.12\mu$m.}
\label{pix_cell}
\end{figure}

By addressing a row and selecting a group of 32 columns for each
conversion cycle, the window of interest inside the ASIC is read out.
Importantly, this readout operation can be done while sampling
continues, providing continuous pipelining and subsequent deadtime
reduction.  This decision to move the high-speed clock and registers
off-chip also means that the size of each storage cell can be
significantly reduced.  A schematic of the base BLAB1 storage cell is
shown in Fig.~\ref{scheme}, where the comparator is simply a
differential NMOS pair.  The corresponding layout is shown in
Fig.~\ref{pix_cell}, where the overall dimensions are $40\lambda$ by
$139\lambda$, where $\lambda = 0.12\mu$m.  This corresponds to 4.8$\mu
$m by 16.68$\mu $m, or about 80$\mu m^2$ required per storage cell.

Therefore the core of the sampling array requires only 5.25$mm^2$ of
chip area, permitting more than an order of magnitude improvement in
storage density compared with existing
devices~\cite{LAB3,ATWD,DRS,Stefan}.  Reducing the cell size and
subsequently the storage capacitance also helps improve the bandwidth
that can be coupled into each storage cell.  Since the ``on''
resistance of the switch is relatively high ($R_{\rm on} \simeq
5k\Omega$), frequencies above

\begin{equation}
f_{\rm 3dB} = {1\over 2\pi R_{\rm on} C_{\rm pix}}
\end{equation}

will roll off for a given pixel capacitance $C_{\rm pix}$.  The
extracted capacitance value for the layout in Fig.~\ref{pix_cell} is
approximately 14 fF.  Therefore the expected $f_{\rm 3dB}$ from the
common input bus line into each storage cell is approximately 2.3
GHz.

We note that the size of the storage cell can be reduced further by
removing individual sample delay timing chains from each storage row.
As seen in the bottom of Fig.~\ref{pix_cell}, this inverter pair is
more than half the area of the storage cell.  For power dissipation
reasons, this removal turns out to be important, as will be discussed later.  


A further benefit of decoupling the latching register and clocking
functionality is that the conversion clock can be run at a much higher
speed inside the FPGA, since it is routinely fabricated in either a
65nm or 90nm process, compared with the relatively coarse 250nm
(0.25$\mu $m) process of BLAB1.  Typically with the chosen Xilinx
Virtex family employed we are able to use a 500MHz clock, and record
the phase of the clock as well, thereby effectively having a 1ns least
count.  Separate testing indicates that this TDC performs very close
to the ideal binary interpolation $1\over \sqrt{12}$ limit ($\sim
300$ps), as reported previously~\cite{JINST}.  Moreover, the number of
bits of resolution or precision can be completely configurable, which
permits a trade-off of the readout latency versus required sample
resolution for various applications.  We note in passing that there is
a potentially much better method based upon applying this same
waveform sampling technique to the timing encoding of the comparator
output.  The TDC least count would then become 1ns $\rightarrow 
170$ps, and fitting the output shape, better than binary encoding time
resolution may be possible.

While the coupling into individual storage cells can support high
analog bandwidth ($\geq 2$GHz), the cumulative capacitance seen when
trying to drive the array of 64k cells is very problematic.  The
extracted capacitance of each of the $2^{16}$ switch drains is about
1.5fF, which sums to a total array capacitance $C_{\rm array}$ of
98pF.  Clearly, for a reasonably low input coupling impedance of
$Z_{\rm in} = 50\Omega$, this bandwidth limitation to

\begin{equation}
f_{\rm 3dB} = {1\over 2\pi Z_{\rm in} C_{\rm array}} = 32.5\; {\rm MHz}
\end{equation}

would be completely unacceptable.  Therefore a 3-level buffer tree has
been employed, to reduce the loading seen at each stage of signal
fan-out.  The unity gain for zero capacitance of these buffer
amplifiers is in excess of 1GHz.  In retrospect, the choice of
fanout: $1 \rightarrow 16 \rightarrow 128$ was not optimal, as the
capacitance of the intermediate state was rather high and limits the
performance, as will be shown in the testing section.

\indent The sampling speed is controlled by adjusting the VDD/VSS
supply voltages of one of the two inverter-inverter delay stages
between each adjacent sampling cell in a particular sampling row.  

As mentioned early, by addressing a row and pulsing the first cell of
that particular row, a write strobe then propagates along the row
until it reaches the last cell in the row.  The leading edge of the
pulse closes the switch and the trailing edge opens the switch, at
which point the analog voltage value is stored.

Upon the determination of an external trigger condition, further
sampling to the row(s) or interest are blocked in firmware and a
ramping voltage is generated by using a constant current source and
reference capacitor, as mentioned earlier.  The ramping voltage for
the BLAB1 can be generated using either an external capacitor or an
on-chip capacitor.  A external capacitor is necessary for slower
ramping speeds.  The current source is set by an external resistor.  A
unique feature of the BLAB1's digitization technique is that the ADC
resolution does not have a default value.  For a fixed clock
frequency, reducing the ramping voltage speed will increase ADC
resolution.  However, by using a slower ramp, it will take longer to
digitize.

\indent BLAB1 was designed to be a low power ADC.  Three voltage
sources are required to operate the BLAB1.  A voltage source of 2.5
volts is the main power source.  An adjustable VDD source is used to
control the sampling speed.  A pedestal voltage, typically 1.3 volts,
is used to set the DC offset of the RF input.  When in quiescent mode,
the power draw can be 10mW or less.  A list of the key BLAB1
specifications are summarized in Table~\ref{specs}.

\begin{table}[hbt]
\caption{\it Important BLAB1 ASIC Specifications.}
\label{specs}
 \begin{center}
    \begin{tabular}{|l|c|} \hline
{\it Item }& {\it Value }  \\ \hline\hline
Sampling Input Channels  & 1  \\ \hline
Storage rows &  128 \\ \hline
Storage cells/row &  512 \\ \hline
Total storage cells &  65,536 \\ \hline
Sampling speed (GSa/s)&  0.1 - 6.0  \\ \hline
Storage record &  10.9 - 655 $\mu$s  \\ \hline
Wilkinson Outputs & 32 \\ \hline
Operation mode & continuous storage/readout \\ \hline 
100ns window readout & 80$\mu $s (5.12 GSa/s, 12-bits) \\ \hline 
Full chip readout & $\sim $10ms (12-bits) \\ \hline \hline
     \end{tabular}
  \end{center}
\end{table}   



\section{Readout Test System}

\begin{figure}[ht]
\vspace*{0mm}
\includegraphics[width=7.6cm]{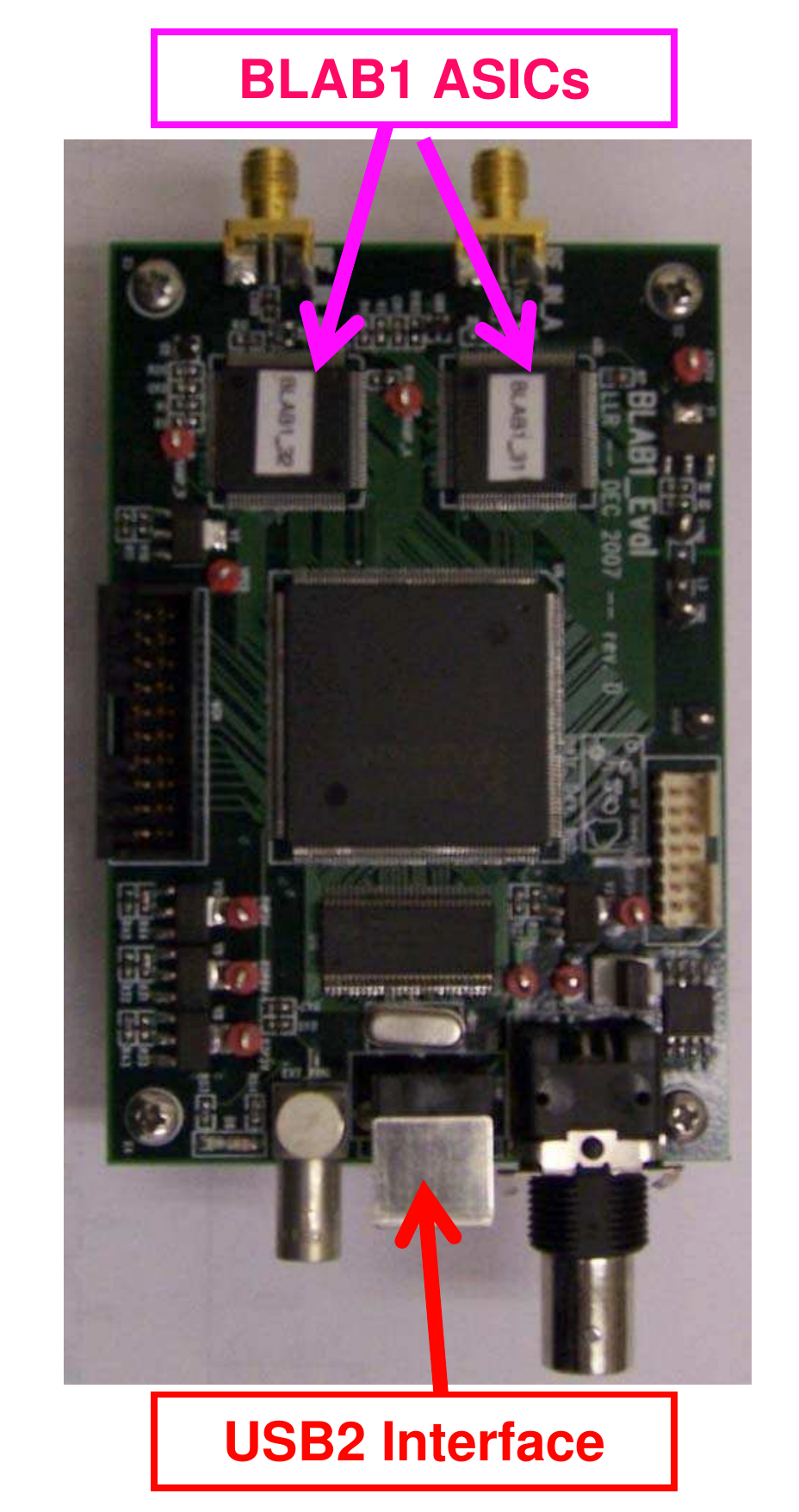}
\vspace*{0mm}
\caption{Photograph of the BLAB1 differential timing performance
evaluation circuit board, with key components indicated.}
\label{BLAB_eval_PCB}
\end{figure}

A series of printed circuit boards have been fabricated to evaluate
various aspects of BLAB1 performance.  Beyond this, these evaluation
devices are proving useful for instrumenting a next generation of Cherenkov
radiation detectors~\cite{PD07}.  A photograph of a 2 BLAB1 ASIC (precision
differential timing evalution) circuit board is shown in Fig.~\ref{BLAB_eval_PCB}.

\begin{figure}[ht]
\vspace*{0mm}
\includegraphics[width=7.6cm]{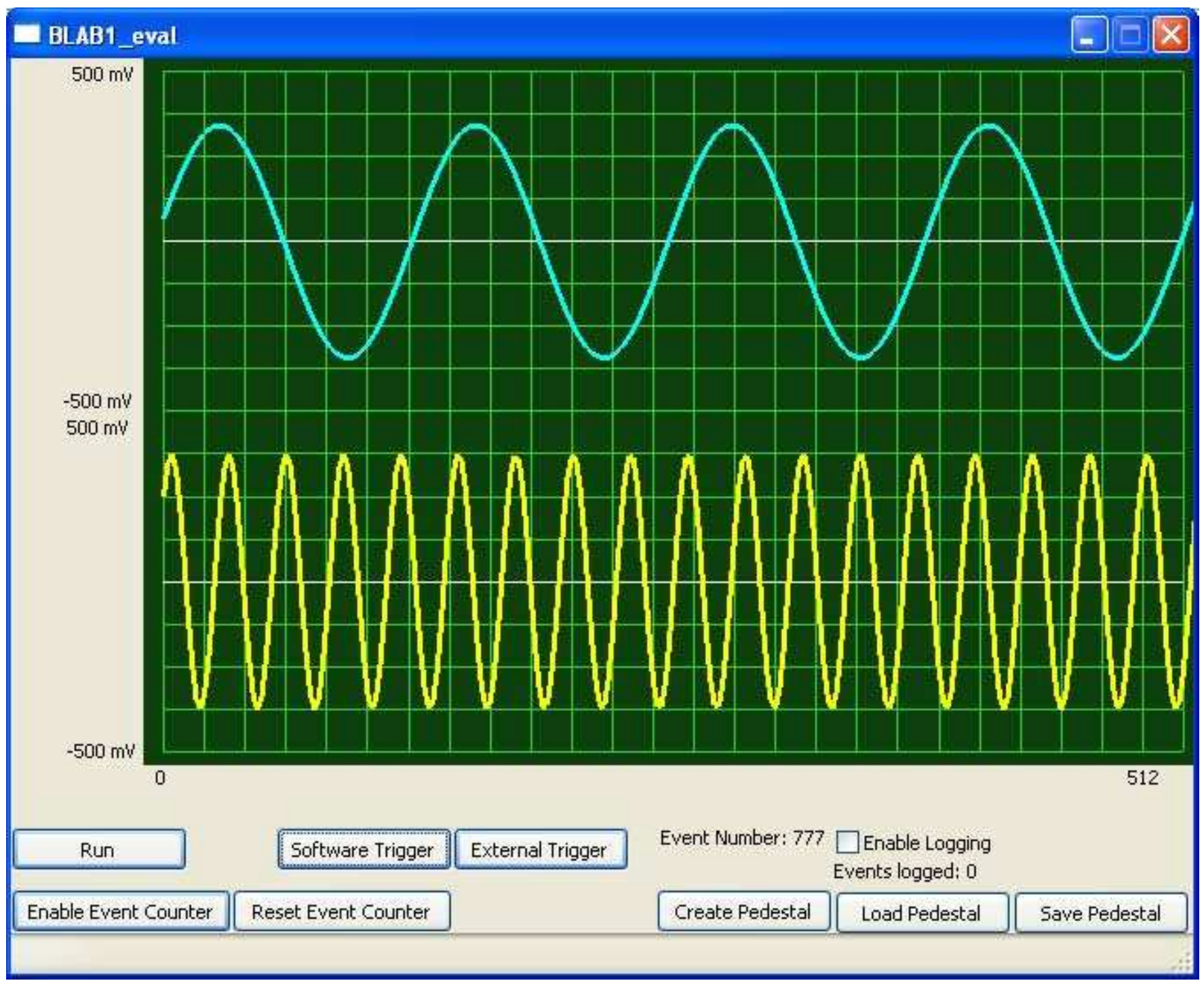}
\vspace*{0mm}
\caption{Screen capture of the acquistion/control program.}
\label{screen_capture}
\end{figure}

The three main components on this circuit board are two BLAB1 chips,
an FPGA (largest package in center), and a Universal Serial Bus (USB)
interface.  The external communication protocol is USB 2.0.  A USB
microcontroller, the Cypress CY7C68013-56PVC, located on the circuit
board interprets the USB 2.0 protocol and controls the flow of data
being sent and received from the FPGA to a computer interface.  The
FPGA used is a Xilinx XC3S400 and controls the digital logic and
timing for the BLAB1 readout.  An internal FPGA RAM buffers the data
while the data is being dumped into the USB data stream.  A custom
readout and control software utility was developed using the wxWidgets
tool kit \cite{wxWidgets}, a screen shot of which is shown in
Fig.~\ref{screen_capture}.  

In this configuration, it becomes apparent that this BLAB1
``oscilloscope on a chip'' can, with this small readout board, turn
any PC (or laptop) into a high-performance digital signal
oscilloscope.  This software package sends commands to the FPGA and
records the BLAB1 data via the USB 2.0 interface.  Running this
utility on a standard PC, a sustained triggered event rate of
approximately 7kHz (single row readout) has been demonstrated.  This
rate should not be considered a hard limit as neither the software nor the
firmware was optimized for speed.  The sampling rate is controlled by
setting a DAC, which then adjusts the VDD voltage (ROVDD) of the
on-chip voltage-controlled delays.



\section{Basic Sampler Performance}

Employing the test system described in the previous section and its
variants, a number of the basic performance parameters of the BLAB1
have been evaluated.  Because timing performance is such a critical
feature of this sampling device, it is described in detail in a
subsequent section.

\subsection{Sampling speed}  

\begin{figure}[ht]
\vspace*{0mm}
\includegraphics[width=7.3cm]{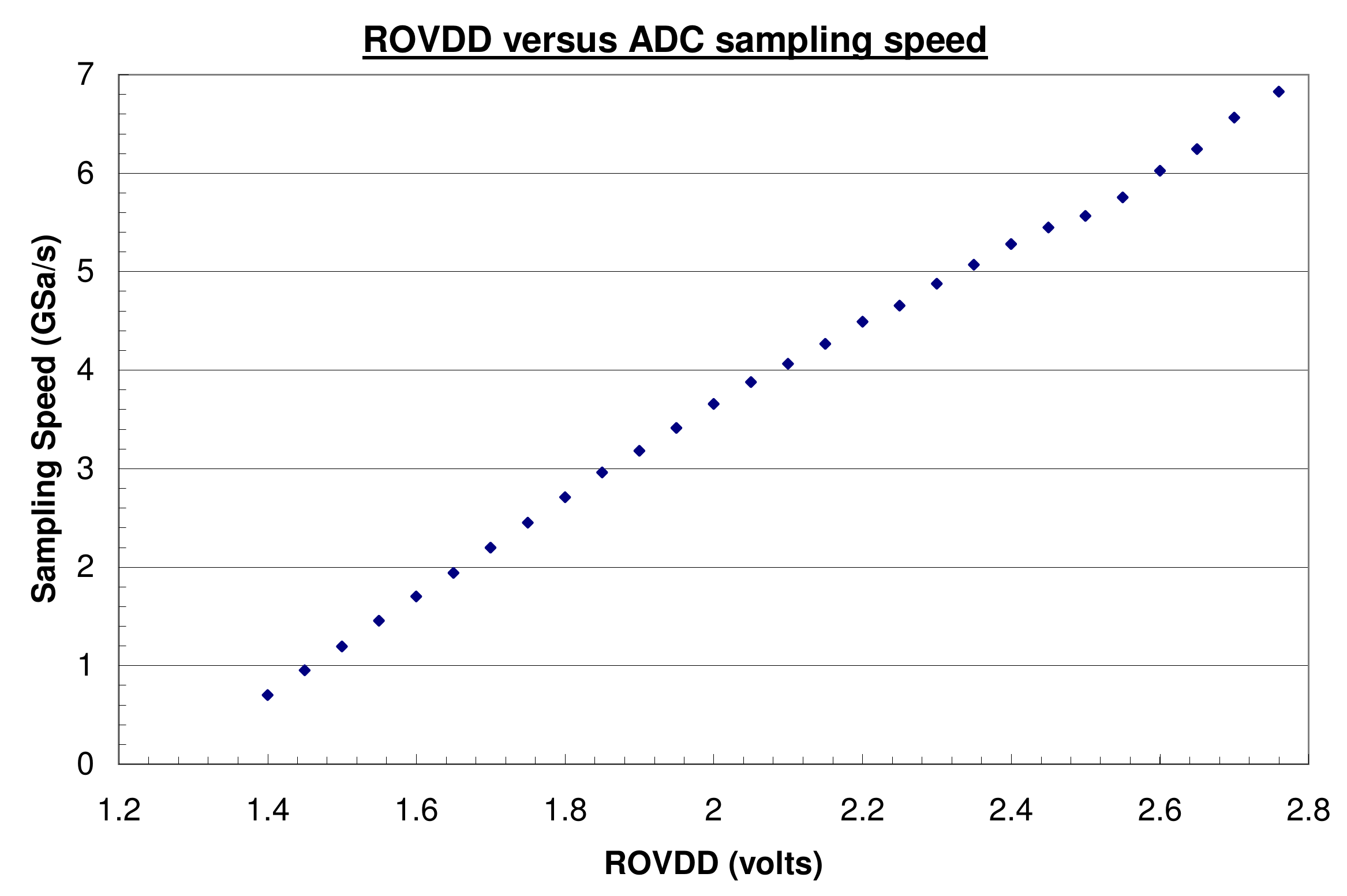}
\vspace*{0mm}
\caption{Sampling rate as a function of the ROVDD control voltage,
where extended operation ($>2.5V$) is possible.}
\label{rovdd}
\end{figure}

Determination of the sampling speed is made by measuring the time
interval between insertion of the timing strobe and appearance of the
output pulse from the last cell of the row, minus pad buffer delays.
The sampling speed is calculated by taking the number of cells in a
row and dividing it by the propagation time for a given control
voltage setting.  A plot of the sampling speed versus control voltage
(ROVDD) is shown in Fig.~\ref{rovdd}, where it is seen that sampling
rates from below 1.0 GSa/s to above 6.0 GSa/s are possible.

\begin{figure}[ht]
\vspace*{0mm}
\includegraphics[width=7.6cm]{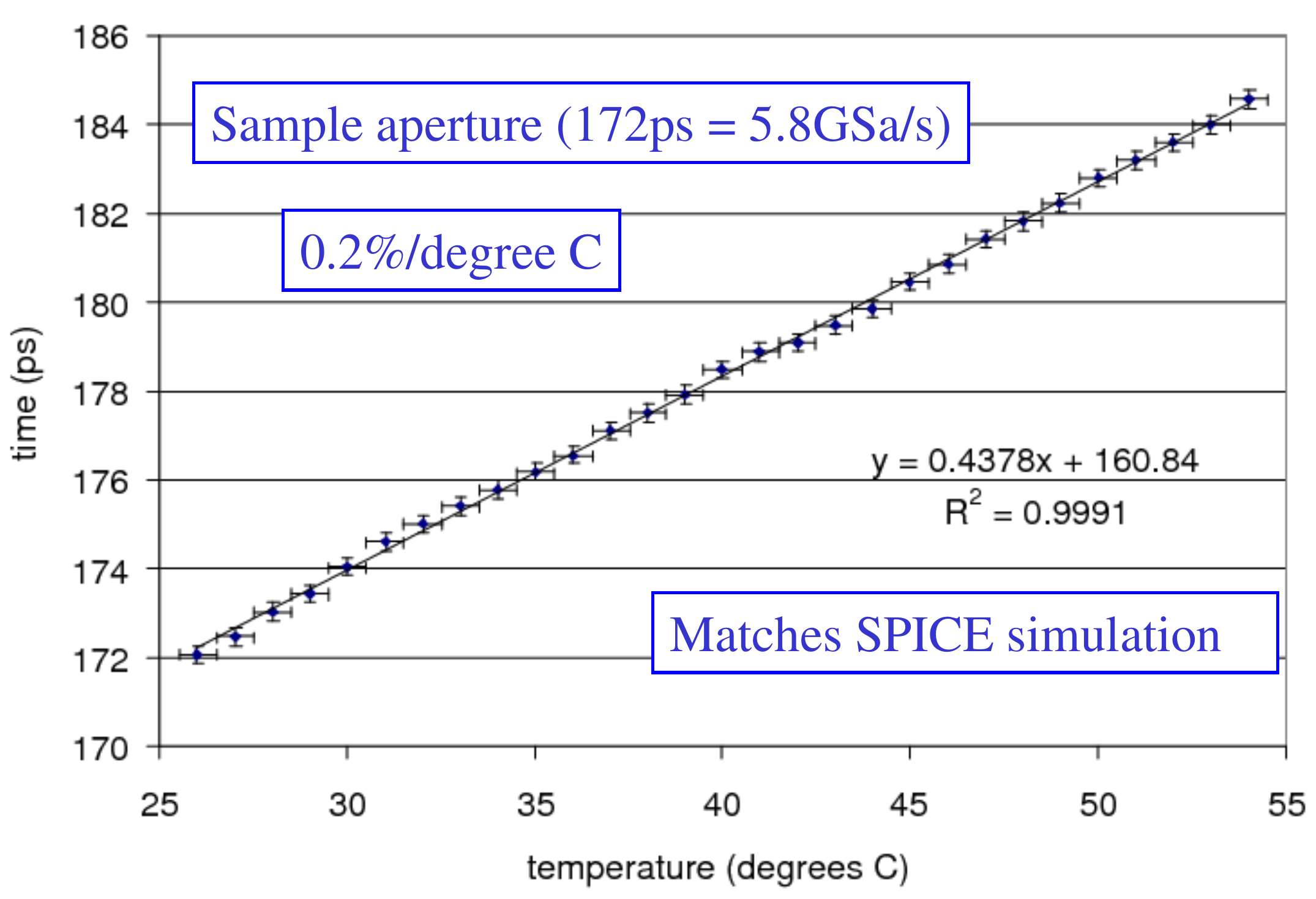}
\vspace*{0mm}
\caption{Temperature dependence of the sampling rate.}
\label{ratedep}
\end{figure}

One potential disadvantage of this voltage controlled delay technique
is that the circuit is temperature dependent.  This dependence is seen
in Fig.~\ref{ratedep} and is roughly 0.2\%/$^{\circ}$C, and completely
matches expectation from SPICE simulation.  While for many
applications this variation would not be significant, and can
potentially be calibrated out with an external reference clock
\cite{LAB3}, the delay can also be monitored and stabilized using a
firmware control loop.

\subsection{Noise performance} 

\begin{figure}[ht]
\vspace*{0mm}
\includegraphics[width=7.6cm]{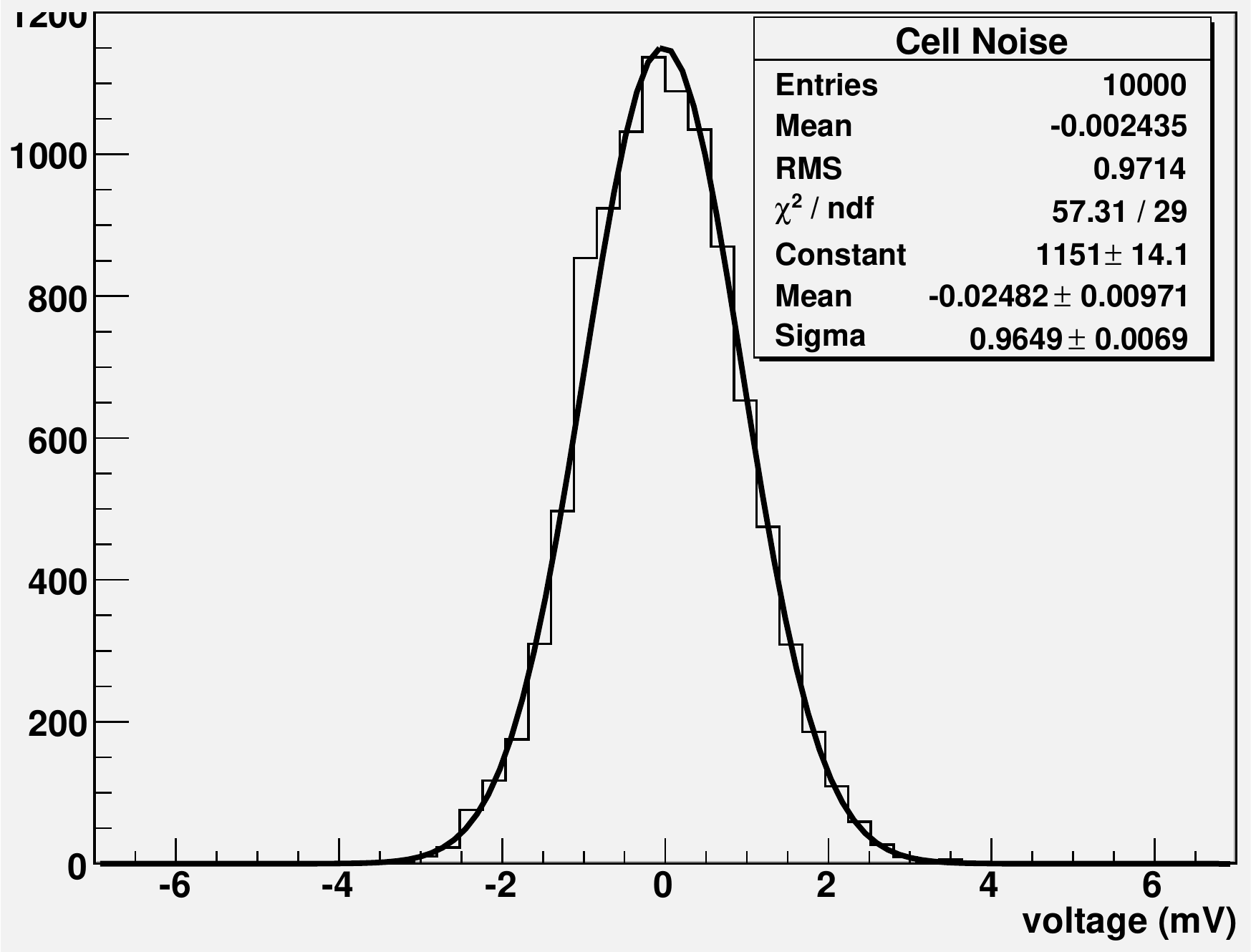}
\vspace*{0mm}
\caption{A representative storage cell noise distribution, where a
Gaussian fit yields a noise level of about 1 mV RMS.}
\label{CELL_NOISE}
\end{figure}

Noise distributions were measured for all storage cells in the process
of determining the pedestal values.  These measurements are made by
terminating the BLAB1 analog input, and reading each cell multiple
times.  An example of the the noise distribution for a typical storage
cell is shown in Fig.~\ref{CELL_NOISE}, which represents the ensemble
mean noise average of about 1 mV RMS.  With an input dynamic range of
greater than 1 Volt (1.5V nom.) and this average noise level, each
stored sample represents 10 real ADC bits of resolution, which is very
competetive with commercially available, large power-dissipation ADCs
\cite{national_semi}.

For comparison, the expected RMS noise due to the small charge quantization 

\begin{equation}
v_{\rm RMS} = \sqrt{ {kT\over C_{\rm pix}}}
\end{equation}

where $k$ is Boltzmann's constant and we take $T$ to be 300K.
Plugging in the $C_{\rm pix}$ from above, we expect a contribution due
to this ``kTC'' noise of

\begin{equation}
\sqrt{ {25.9m[e{\rm V}]\cdot 1.6\times10^{-19}\left[ {{\rm C}\over e} \right] } \over
14{\rm f\left[ {\rm {C\over V}} \right]} } = 0.54{\rm mV}
\end{equation}

which subtracted in quadrature indicates that the excess ASIC and
board level noise is approximately 0.84mV, and could perhaps be
improved through better layout.

\subsection{Analog bandwidth}

\begin{figure}[ht]
\vspace*{0mm}
\includegraphics[width=7.6cm]{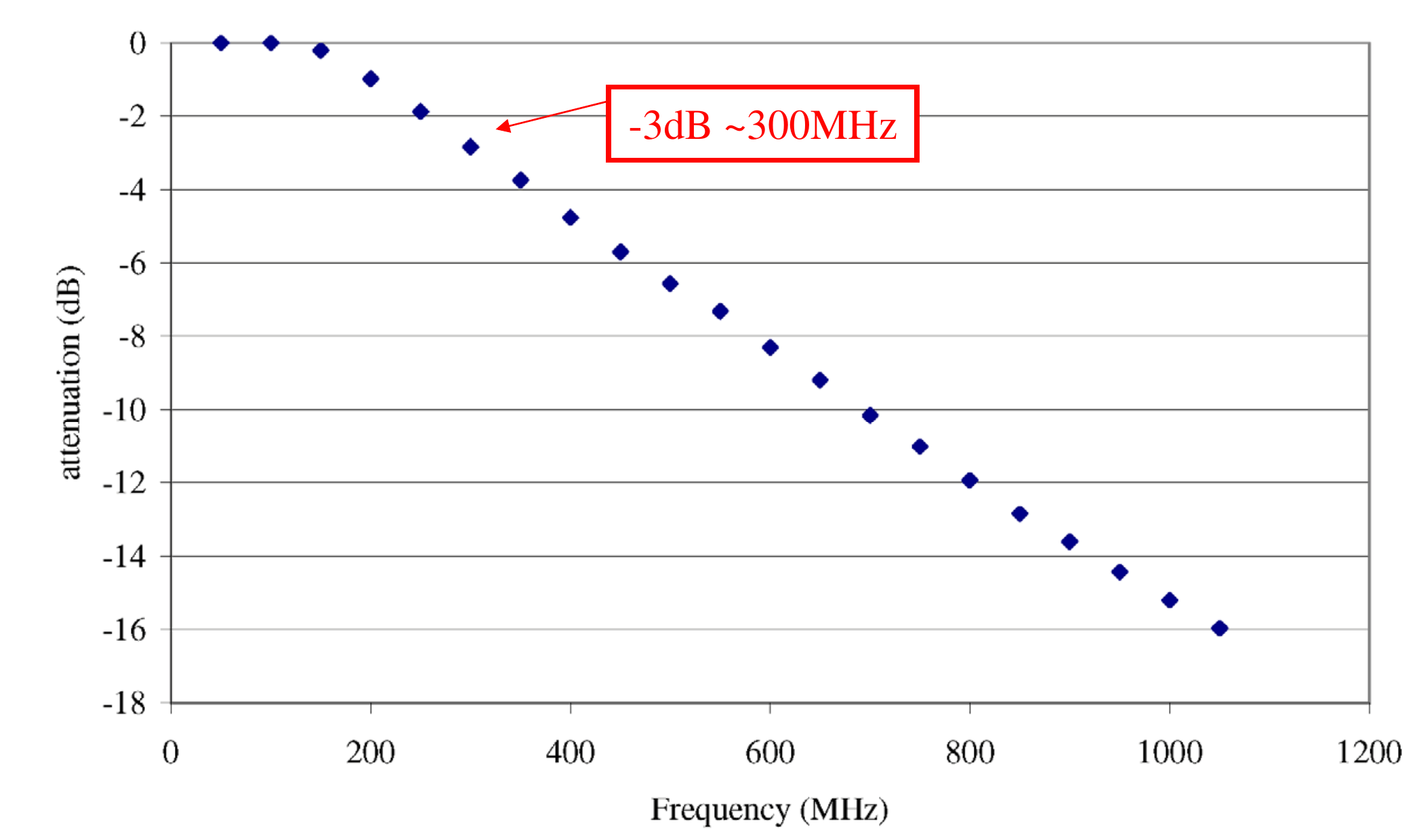}
\vspace*{0mm}
\caption{Measured frequency response of the BLAB1.}
\label{RF_LOSS}
\end{figure}

A determination of the analog frequency response of the BLAB1 ASIC was
performed by recording fixed amplitude sine waves of varying
frequencies and comparing the ratio of the actual amplitude to the
recorded amplitude.  The amplitude roll-off versus frequency is shown
in Fig.~\ref{RF_LOSS}, where the -3 dB attenuation point is about 300
MHz, and the -10dB point extends beyond 600 MHz.

\begin{figure}[ht]
\vspace*{0mm}
\includegraphics[width=7.6cm]{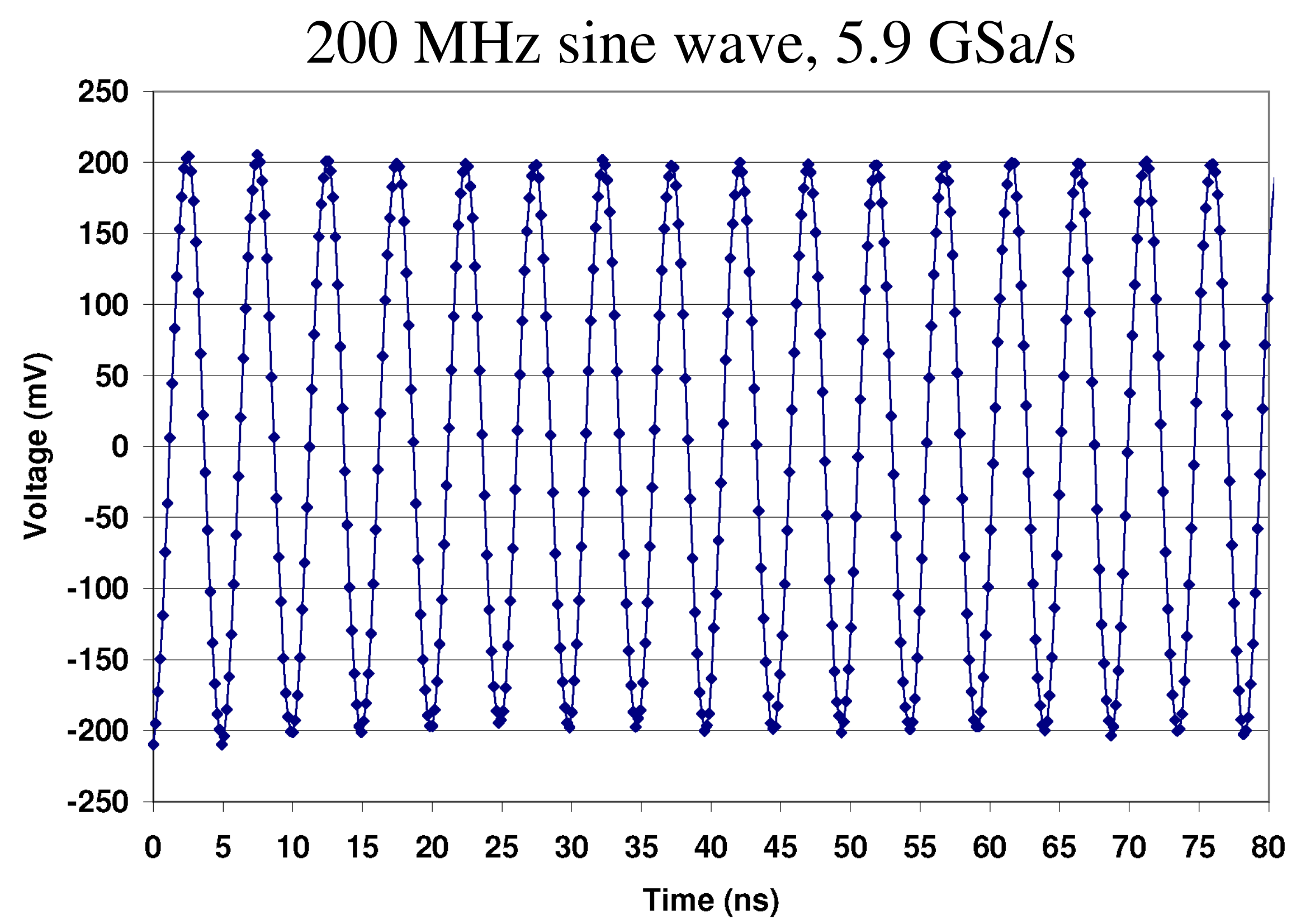}
\vspace*{0mm}
\caption{Reference 125MHz sine wave recorded with the BLAB1.}
\label{RF_sine}
\end{figure}

To illustrate how this performance corresponds into the ability to
sample an RF sine wave, consider the uncalibrated waveform of
Fig.~\ref{RF_sine}.  As the noise is small, deviations from a smooth
curve give an indication of the level of calibration required in the
following precision timing section.

\subsection{Leakage current} 

\begin{figure}[ht]
\vspace*{0mm}
\includegraphics[width=7.2cm]{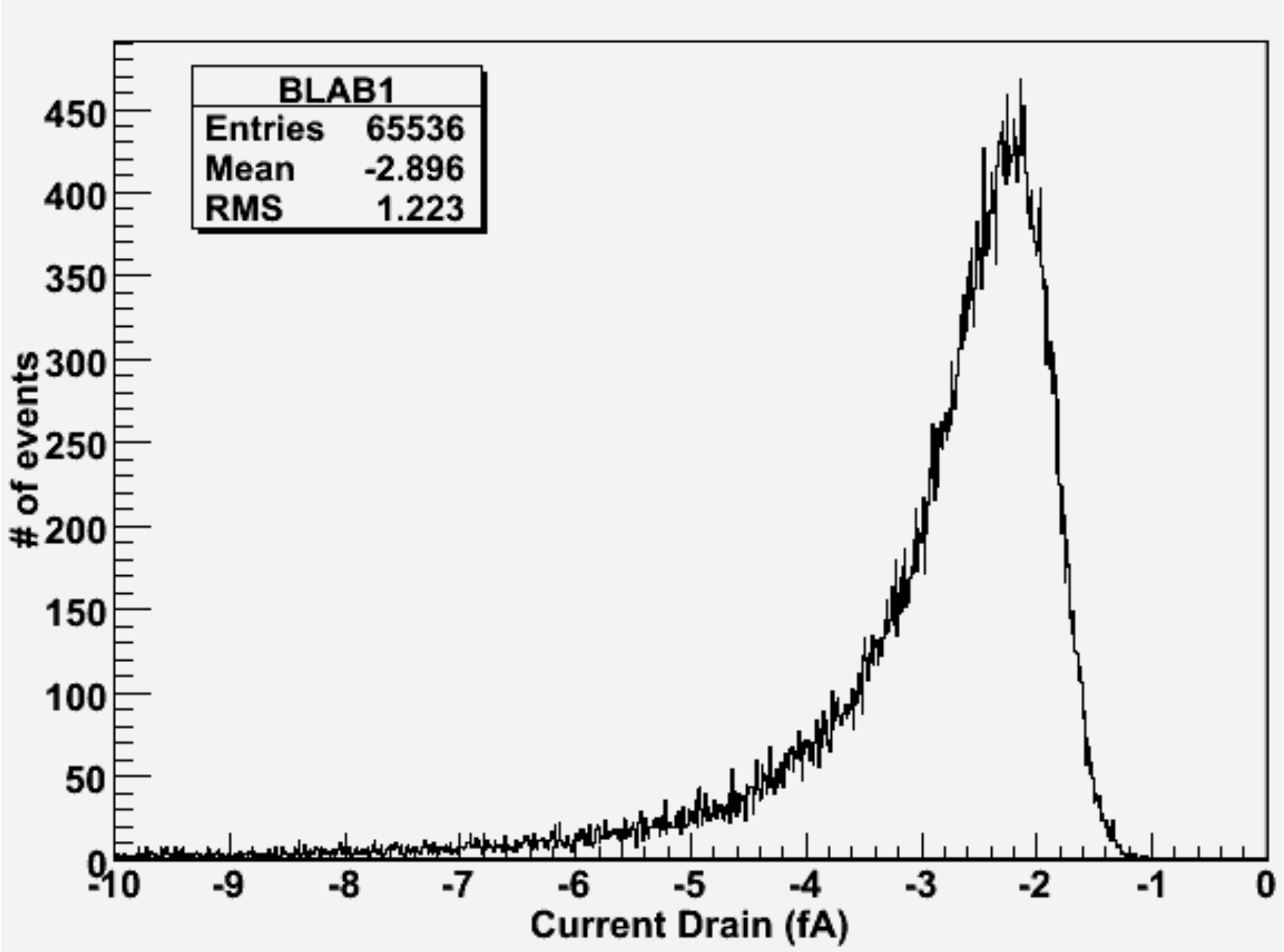}
\vspace*{0mm}
\caption{Leakage current histogram for all cells of a given BLAB1
device, where the mean leakage current is slightly under 3fA.  }
\label{DRAIN_1D}
\end{figure}

Because leakage current is a concern for long storage times, and the
array contains a large number of samples, which potentially take a
long period to read out completely, this issue was studied
extensively.  A measurement of the leakage current for all 64k
sampling capacitors was performed.  This measurement was done by
terminating the BLAB1 analog input and reading out each cell
repeatedly, without a write update, for 20 seconds.  A summary histogram
of the leakage current determined for all storage cells from a fit to
each leakage current slope is plotted in Fig.~\ref{DRAIN_1D}.

For reference, these values are in quite good agreement with leakage
currents measured previously by our group for a similar TSMC CMOS
process in different fabrication runs \cite{CAP1,CAP2}.  If the effect
of this leakage current is to be reduced to a level comparable with
the noise, the following condition must be met:

\begin{equation}
1mV = \Delta V = {\Delta Q \over C_{\rm pix}} = {{I_{\rm leak} \cdot
\Delta T} \over C_{\rm pix}}
\end{equation}

where $\Delta T$ is the maximum storage to readout interval and
$C_{\rm pix}$ is the pixel storage capacitance.  Using a leakage
current of $I_{\rm leak}$ = 25fA, which is conservatively larger than
almost all storage cells, the maximum readout latency is thus

\begin{equation}
\Delta T = {1{\rm mV} \cdot 14{\rm fF} \over 25 {\rm fC/s}} = 560\mu {\rm s}
\end{equation}

and is discussed in the {\it readout speed} subsection next.  In
general deep storage is needed for trigger latency buffering and a
far smaller window of interest need only be read out.

\begin{figure}[ht]
\vspace*{0mm}
\includegraphics[width=7.6cm]{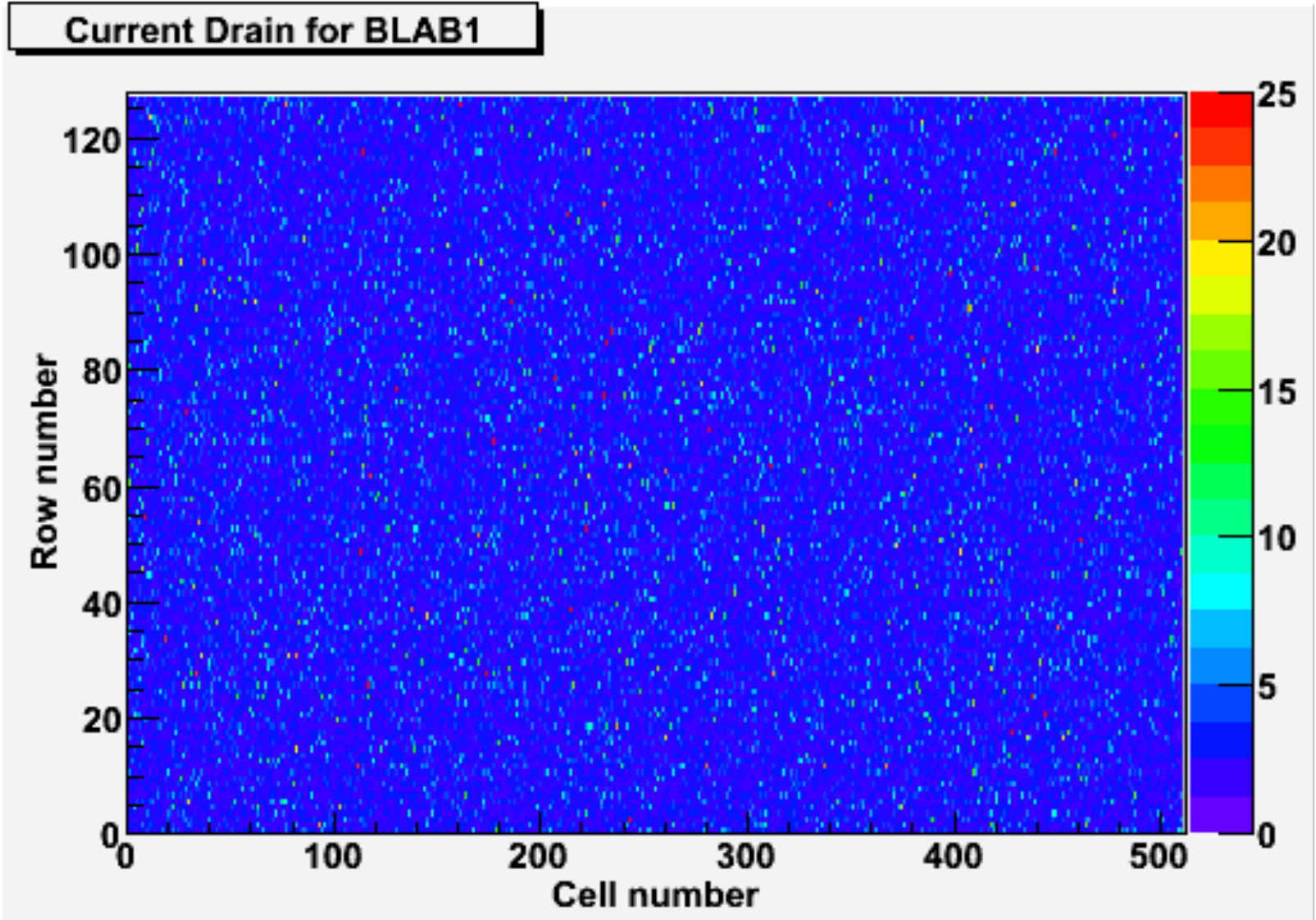}
\vspace*{0mm}
\caption{Array summary plot of the leakage current for all 64k pixels,
where vertical is row number, horizontal is sample number and color
code is in units of femto-Amperes.  No pattern is observed in the
layout.}
\label{DRAIN_2D}
\end{figure}

It had been posited that the more extreme leakage current values might
correlate with the co-location of other logic or structures at the
sampling array periphery.  This conjecture is tested and rejected in
Fig.~\ref{DRAIN_2D}, where the measured leakage current for each cell
is plotted by array location.  No obvious pattern is seen, and the
values are consistent with being completely random.

\subsection{Readout Speed}

As mentioned earlier, there is flexibility in choice of the resolution
versus speed trade-off.  If determined to read out the entire array,
the conversion cycle duration may be expressed as 

\begin{equation}
T_{\rm cycle} = T_{\rm conv} + T_{\rm switch} 
\end{equation}

where $T_{\rm switch}$ is the fixed latency (typically 50ns)
associated with resetting the voltage ramp/changing addresses and
$T_{\rm conv}$ is the interval required for the conversion to n-bits,
given by the expression

\begin{equation}
T_{\rm conv} = 2^n \cdot (1 \; {\rm ns})
\end{equation}

for the 500MHz, dual-phase clock reference used in our measurements.

\begin{figure}[ht]
\vspace*{0mm}
\includegraphics[width=7.6cm]{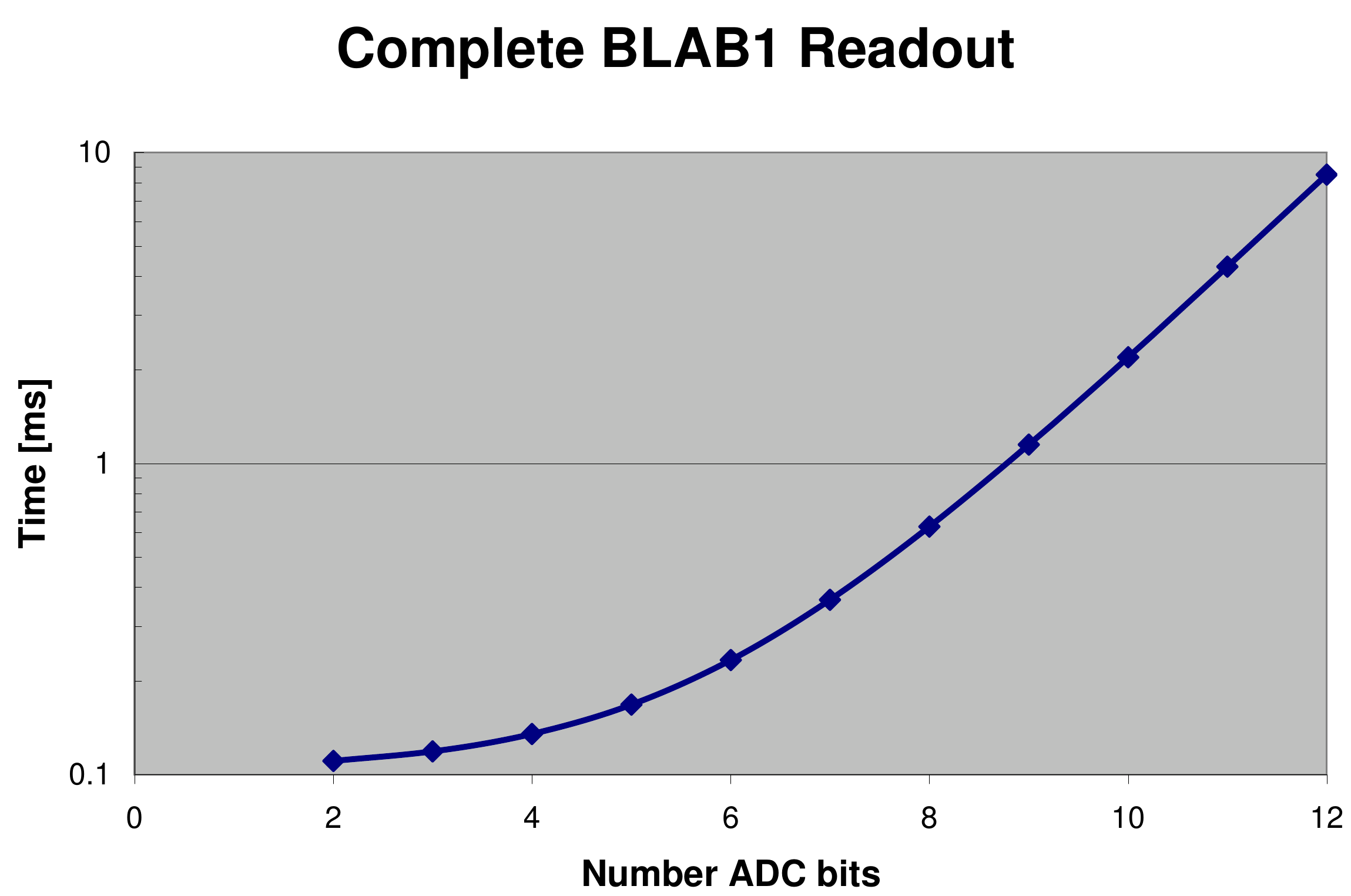}
\vspace*{0mm}
\caption{Time required to read out the entire BLAB1 ASIC as a function
of the number of bits of resolution.}
\label{Conv_time}
\end{figure}

As mentioned previously, full chip readout is a rather extreme case.  For a
detector of the size of a typical high energy physics experiment, for
``fast timing'' signals, something like 100ns is the largest window 
required.  Even for a multi-km scale radio neutrino detector,
the aperture of interest would still only be in the $\mu $s range,
corresponding to less than 10\% of the array, and for which the
readout latency would be less than a millisecond.  For a 100Hz radio
trigger, or a 30kHz collision trigger, the deadtime is negligible for
pipelined operation.

\subsection{Power Dissipation}

During sampling, the power dissipation can be as low as 

\begin{equation}
P = I\cdot \Delta V = {\delta Q\over \delta t}\cdot (2.5{\rm V}) \simeq 15{\rm mW}
\end{equation}

where $\delta Q$ is the inverter transition charge and $\delta t$ is
86ps at the nominal 5.8GSa/s sampling.  During sampling all of the
other biases may be disabled.

Quite unexpectedly, it was observed that lowering $\Delta V $ in the
delay chain (running more slowly) dissipated more power, opposite of
what the expression above would indicate.  Below 2V, significantly
more power was drawn.  Returning to SPICE, it was found that indeed as
the ROVDD is lowered, the leakage current of the inverters becomes
important.  In particular because of the decision to give each storage
cell its own inverter pair.  That $\sim 6\times 10^{4}$ multiplier proved
to be a huge factor and precluded sustained low-speed sampling due to
enormous power dissipation.  Data and simulation agree qualitatively,
though at large current draws it is likely the voltage drop in the
finite resistance of the die power wiring becomes important (and
ignored in simulation).

\begin{figure}[ht]
\vspace*{0mm}
\includegraphics[width=7.6cm]{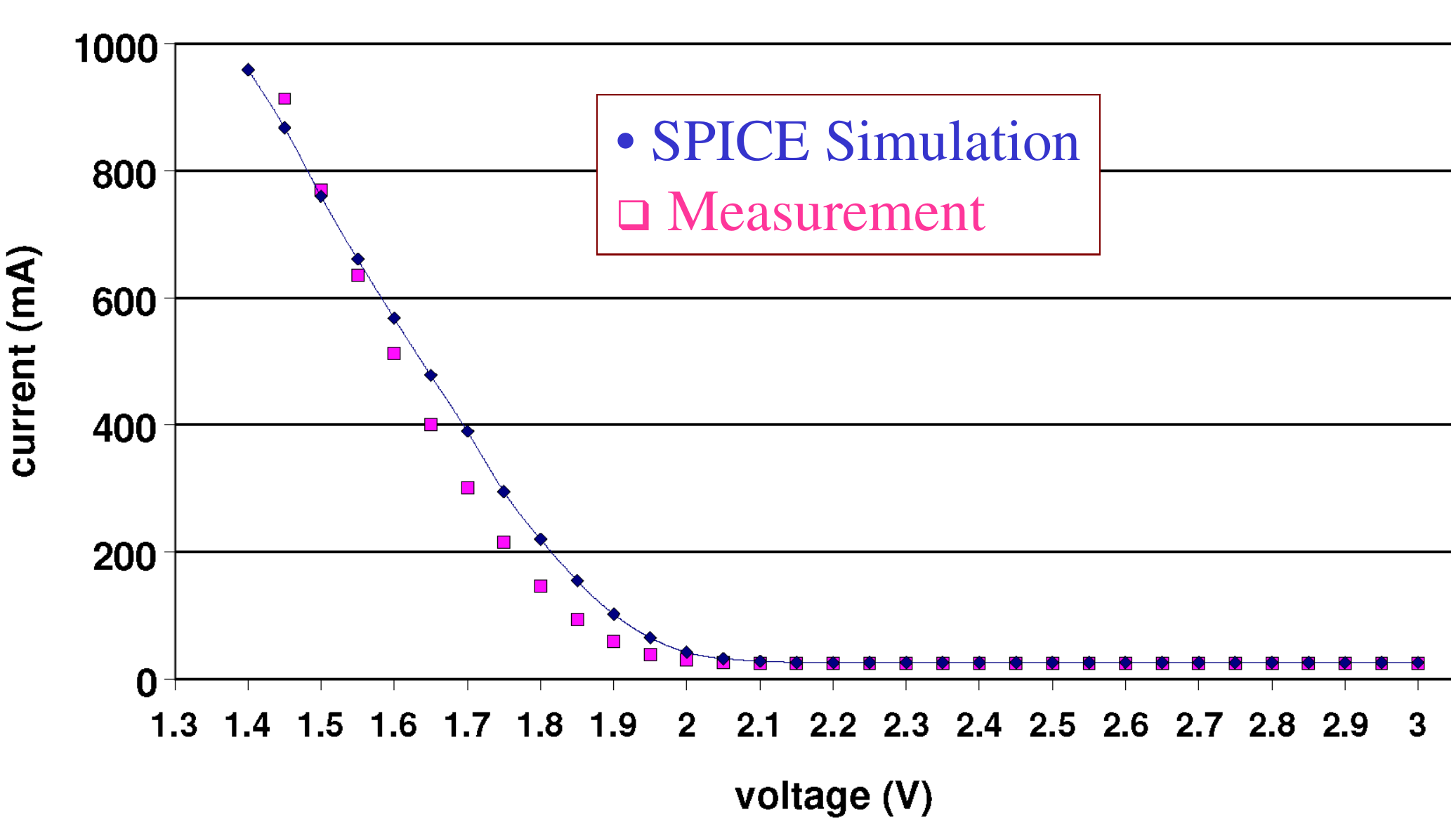}
\vspace*{0mm}
\caption{Simulated versus measured current draw of the entire array of 
voltage-controlled delay inverters.}
\label{power_diss}
\end{figure}

During readout, the current surges and the die subsequently heats
substantially due to an oversight in the original design.  While the
comparator bias currents can be shut down during sampling, when
conversion is required, they must all be operated.  Again a large
multiplier (64k comparators) applies, and even a $10\mu$A comparator
bias leads to a 0.65A surge.  This is addressed in future designs.

\subsection{Concurrent Operation}

\begin{figure}[ht]
\vspace*{0mm}
\centerline{\includegraphics[width=4.0cm]{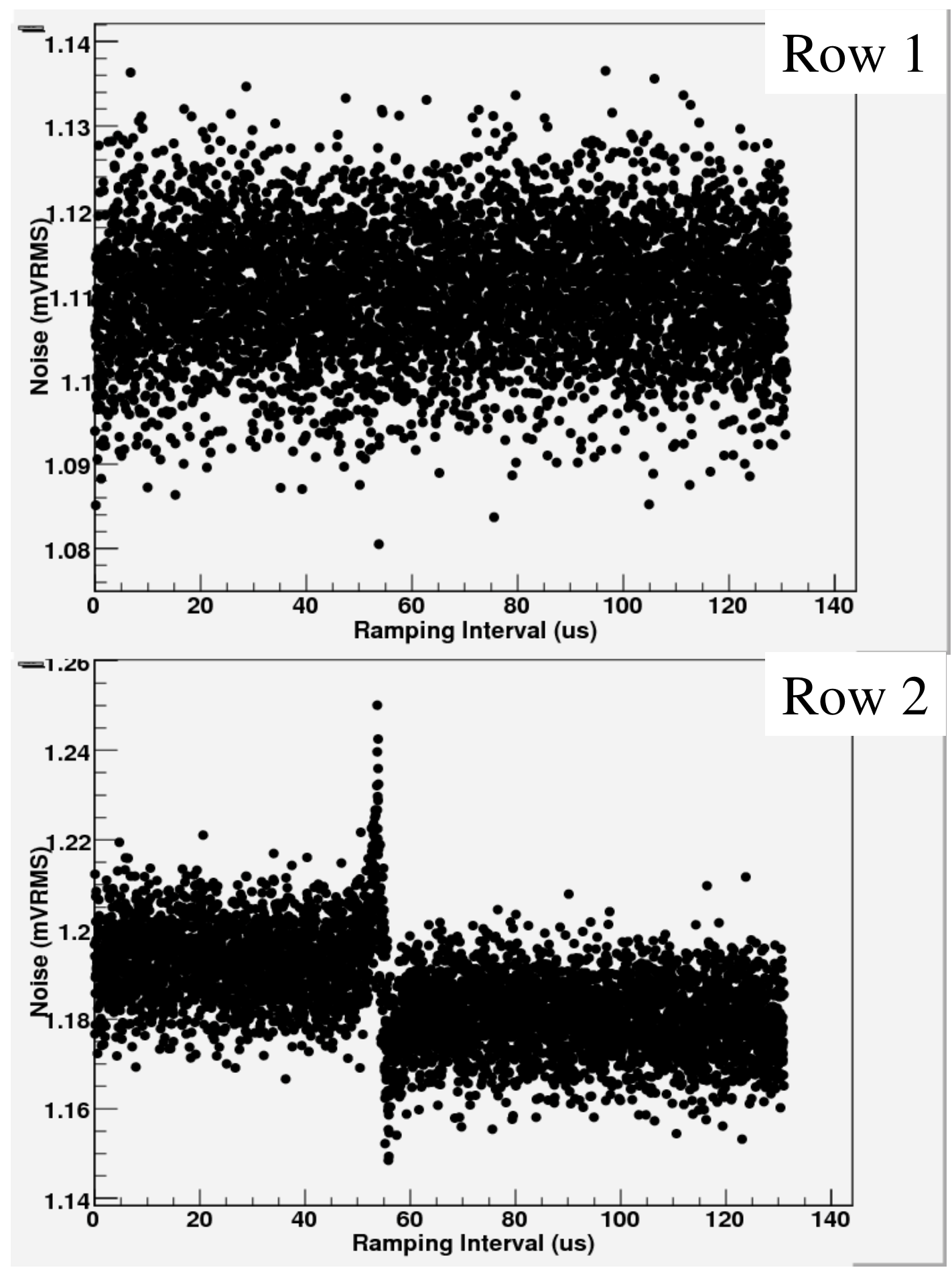}}
\vspace*{0mm}
\caption{Observed noise in storage Channel 2 when
simultaneous readout is performed in Channel 1.}
\label{concur_ops}
\end{figure}

A key feature of the BLAB1 architecture is the ability to operate in a
multi-hit buffer mode, to effectively reduce the deadtime to
negligible levels.  Concurrent readout while continuing to sample can
have a deleterious impact on the quality of storage samples.
Therefore we have performed a noise scan where the delay time of
storage in Row 2 (adjacent row) is varied while Row 1 recording
continues. The result appears in Fig.~\ref{concur_ops}, where a small
amount of cross-talk is observed right about the comparator transition
time for Row 1.  The effect is tiny ($\simeq$1 mV) and can be
neglected.

As an example of the potential benefit, for a future 16 channel BLAB2
ASIC, where a 32ns window (320 samples at 10GSa/s) is recorded from each channel
upon receipt of a Level 1 trigger, the net conversion time to 10 bits
is roughly 160$\mu$s if all these samples are read out.  However, with an
expected hit occupancy in the window for each ASIC (monitored by
trigger out signal) of about 3.2\%, the mean latency for readout is
5.12$\mu $s.  For a 30kHz maximum trigger rate, this is a 15.4\%
deadtime, though with large fluctuations.  Having an 8 deep hold
buffer for each channel (100ns wide), the probability of an overflow
becomes a negligible $5\times 10^{-8}$.



\section{Precision Timing Performance}

Recent developments in high-density, high precision timing
photodetectors are finding applications in Cherenkov detection
techniques for particle identification, as well as medical imaging
applications.  To fully exploit the potential of these devices, robust
performance, fine resolution timing and highly integrated readout
electronics are needed.  Over the decades a number of electronics
techniques have been explored to maximize the timing performance of
photodetector signals.  These include Constant Fraction
Discrimination, multi-level thresholding, charge integration for
threshold timewalk correction, among a long list too extensive to
adequately summarize here.

However, all of these techniques suffer from a number of practical
limitations in actual application, which has served to degrade the
realized performance.  In the end, one simply cannot do better than
having a high-fidelity ``oscilloscope on a chip'' for every sensor
channel.  Cost and data volume precluded this type of waveform
recording until recent generations of SCA ASICs \cite{LAB3,ATWD,DRS}
demonstrated such techniques were practical, especially for large
systems.

We present here some preliminary results of timing resolution tests
with this BLAB1 ASIC.  As these devices are distributed to interested
users around the world, and more clever algorithms for improved timing
performance are considered, further improvements on already promising
results may be obtained.

\subsection{Calibration}

In order to address bin-by-bin timing width differences, a couple of
different calibration techniques have been tried.  The first utilizes
a sine wave zero-crossing technique used for calibrating the LAB3
ASIC\cite{LAB3}.  That technique works best when the frequency of the
sine wave is such that the measured interval between zero crossings
can be uniquely assigned to a limited number of bins between
successive crossings.  
\begin{figure}[ht]
\vspace*{0mm}
\includegraphics[width=5.5cm]{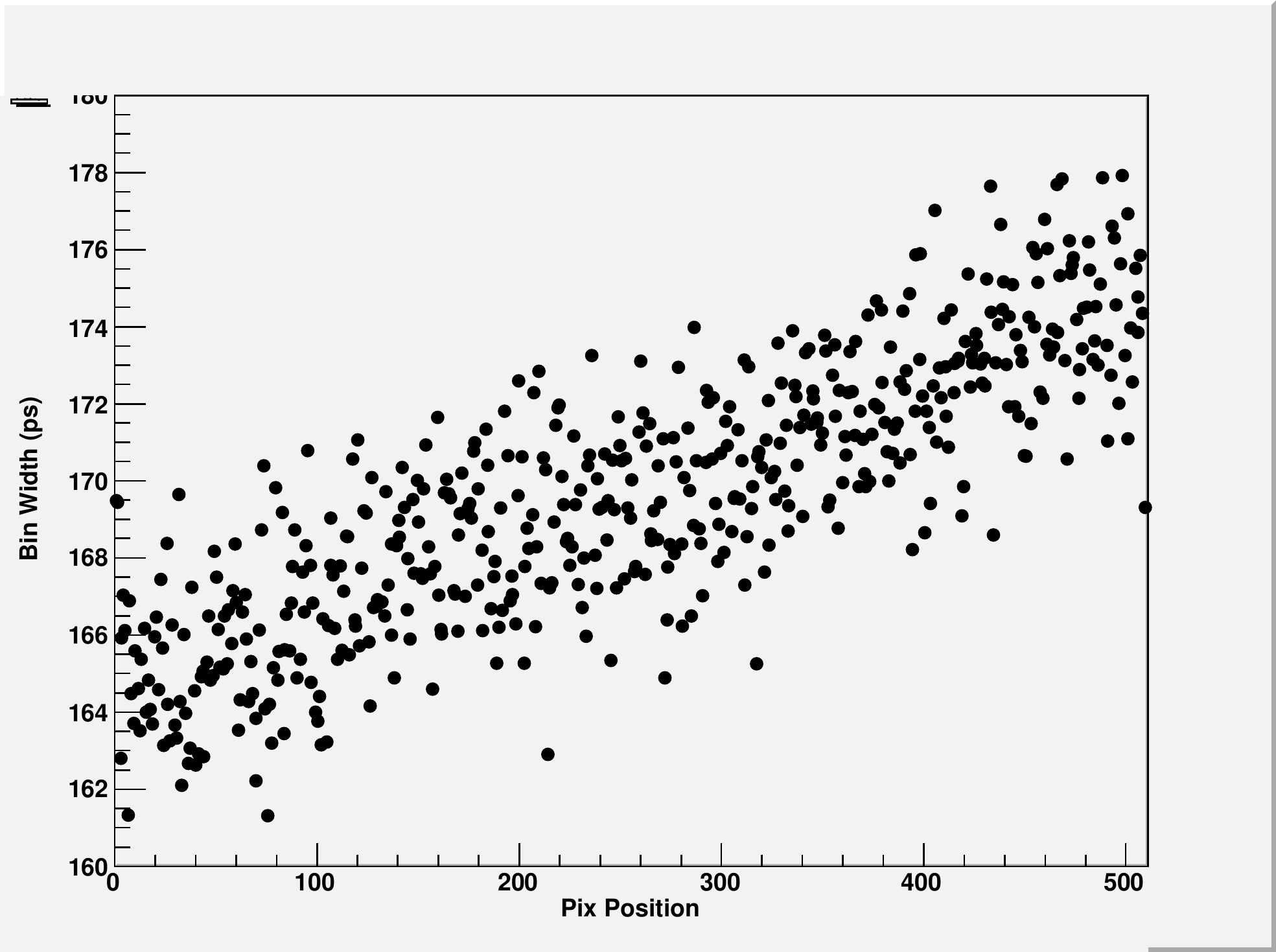}
\vspace*{0mm}
\caption{Residual bin-by-bin sample timing aperture deviations from a
nominal bin width using the histogram occupancy technique described in the
text.}
\label{Occup}
\end{figure}

\begin{figure}[ht]
\vspace*{0mm}
\includegraphics[width=7.6cm]{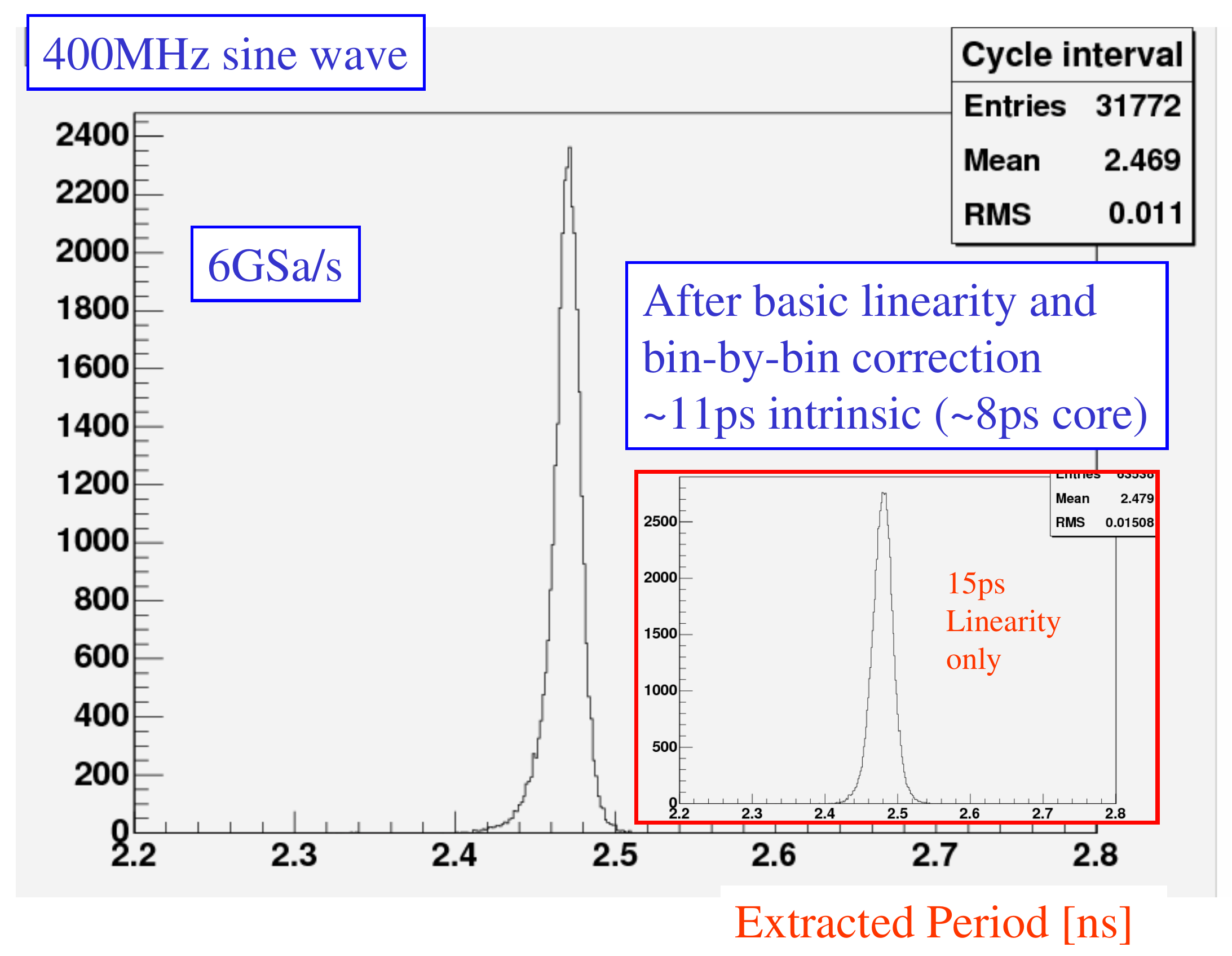}
\vspace*{0mm}
\caption{Results for extracting the zero-crossing timing of a 400MHz
sine wave after the application of the histogram occupancy timing
corrections.  Inset is the result for a simple linear (slope)
correction, and the main plot after a bin-by-bin correction.}
\label{sine_fit}
\end{figure}

Due to intrinsic curvature limitations, this technique has an
irreducible systematic error that is a function of sample rate.  A
more successful technique is to histogram the zero crossings of a sine
wave and use the bin occupancy to derive the effective aperture width,
the residual distribution is shown in Fig.~\ref{Occup}.
The most striking aspect of this distribution is the linear slope
across the array.  Applying only this linear slope correction leads to
the 15ps RMS jitter in the determination of zero crossings for a
subsequent sine wave data set, as seen in the inset distribution in
Fig.~\ref{sine_fit}.  Applying a full bin-by-bin correction improves
the distribution to 11ps RMS, with about an 8ps core.

\subsection{Bench Test Signals}

\begin{figure}[ht]
\vspace*{0mm}
\includegraphics[width=7.6cm]{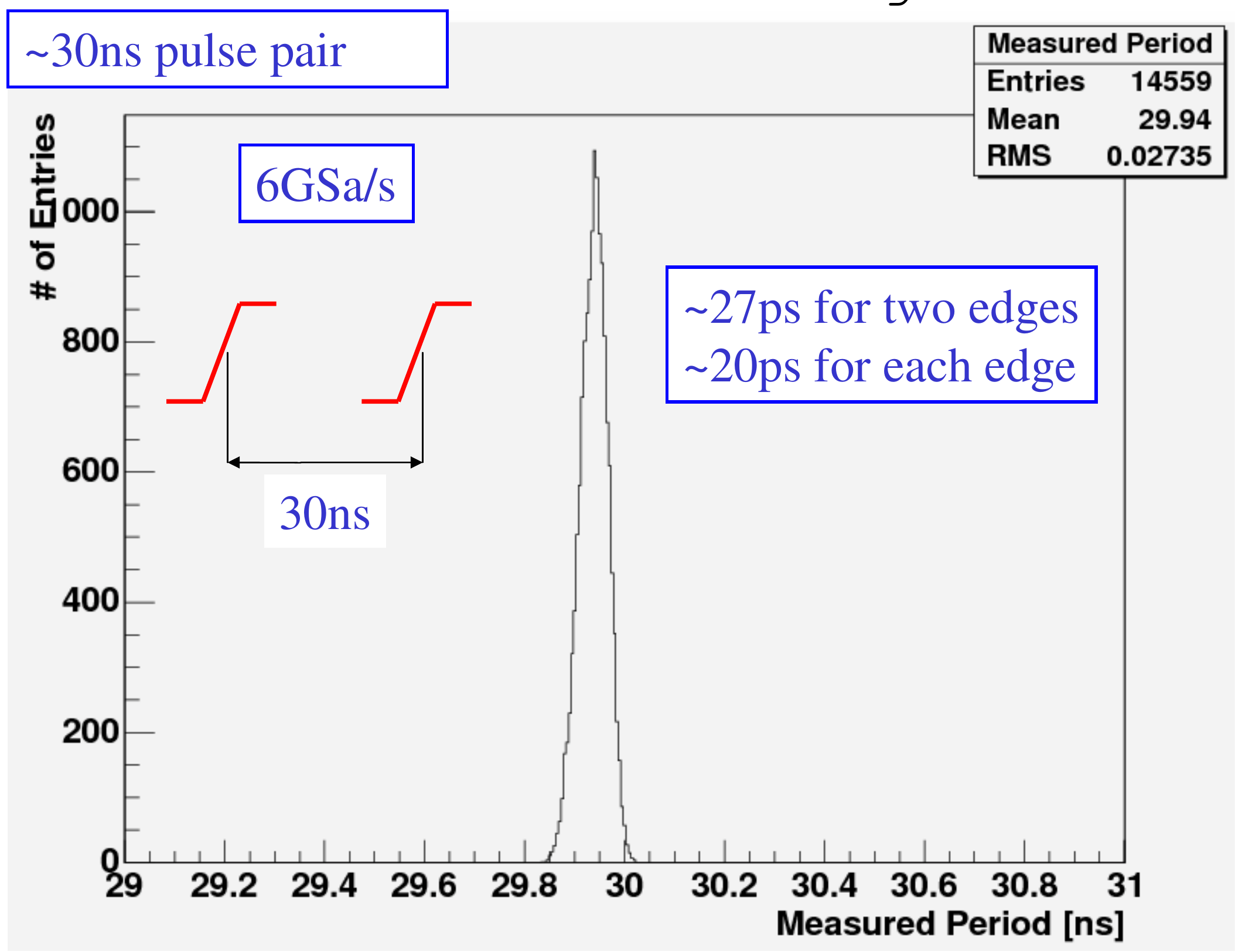}
\vspace*{0mm}
\caption{Timing resolution for a pair of pulses separated by
approximately 30ns.  Each edge can be inferred to be extracted a
factor of $\sqrt{2}$ better.}
\label{pulse_pair}
\end{figure}

Timing performance was then evaluated using a pair of pulses separated
by approximately 30ns.  As seen in Fig.~\ref{pulse_pair}, over this
longer timebase separation, a differential error of 27ps is obtained.
The contribution of each edge then is then estimated as
27ps/$\sqrt{2}$, or 20ps per recorded edge.

For complex curvature along the leading edge of the signal, the timing
resolution obtained is seen to be rather sensitive to the method
choosen to characterize the signal ``hit'' time.  Unless the
photodetector signal is for a single p.e. quanta, the actual shape can
be rather complex and dependent upon photon arrival statistics.  Even
in this simple case, noise and aperture systematics upon the leading
edge can be important and can also be reduced by using multiple
samples to fit to an analytic signal shape.  In general, the estimate
error can improve as something like 1/$\sqrt{N}$ for N samples along
the leading edge.  This is perhaps the most powerful aspect of having
the full waveform samples to fit.  Individual sampling errors can be
averaged out.  Examples are provided in the following subsection,
where it is clear that at the sampling rates being studied, this
waveform recording technique logs many samples on the leading edge,
which can be used to improve the signal timing extraction.

\subsection{PMT signal observation}

\begin{figure}[ht]
\vspace*{0mm}
\includegraphics[width=7.6cm]{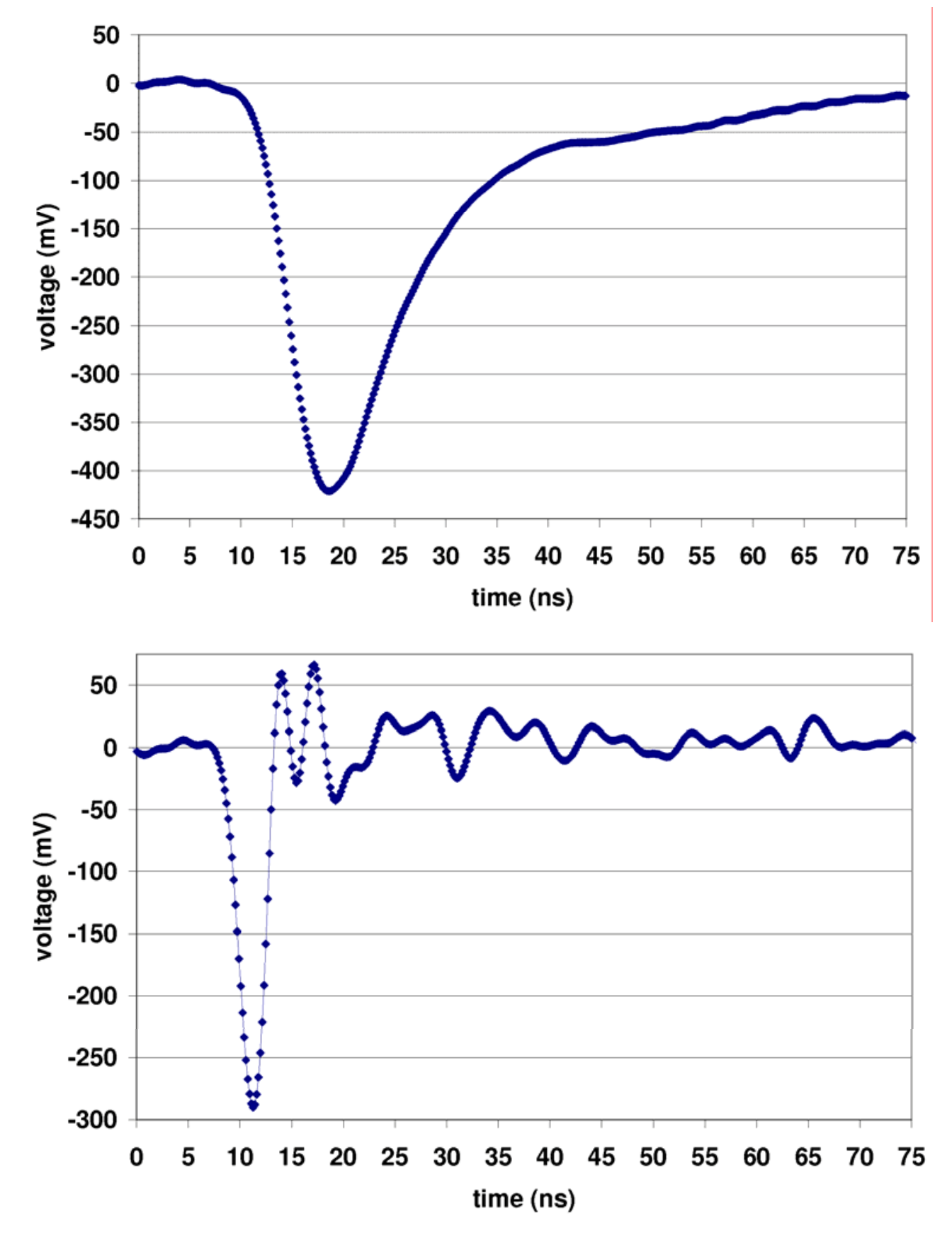}
\vspace*{0mm}
\caption{Example waveforms recorded with a Hamamatsu R6680 fine-mesh
PMT (top) and Burle 85011 Micro-Channel Plate PMT (bottom).}
\label{PMT_plots}
\end{figure}

A convenient feature of the BLAB1 ASIC is that a PMT output
transmitted over a 50$\Omega$ coaxial cable can be directly connected
to the BLAB1 input, as per the diagram of Fig.~\ref{BLOCK}.  Two
example photodetector outputs, intended for fast-timing applications,
are recorded in Fig.~\ref{PMT_plots}.

Both photodetectors specialize in fine time resolution and a direct
comparison is informative.  In the upper figure, the observed signal
is an aggragate of a number of scintillation photons collected from a
bar scintillator described in the next subsection.  At bottom is the
risetime of Micro-Channel Plate photodetector (MPC-PMT), intended for
precise single photon detection. For future sub-10ps devices, the
transit-time spread in the single p.e. amplification process may
limit the ultimate resolution.  

Finally affordable fast electronics may be able, on a
channel-by-channel basis, to measure systematic variations and provide
the requisite compensating corrections to achieve the penultimate
resolution.

\subsection{Belle TOF Counter}

\begin{figure}[ht]
\vspace*{0mm}
\includegraphics[width=7.6cm]{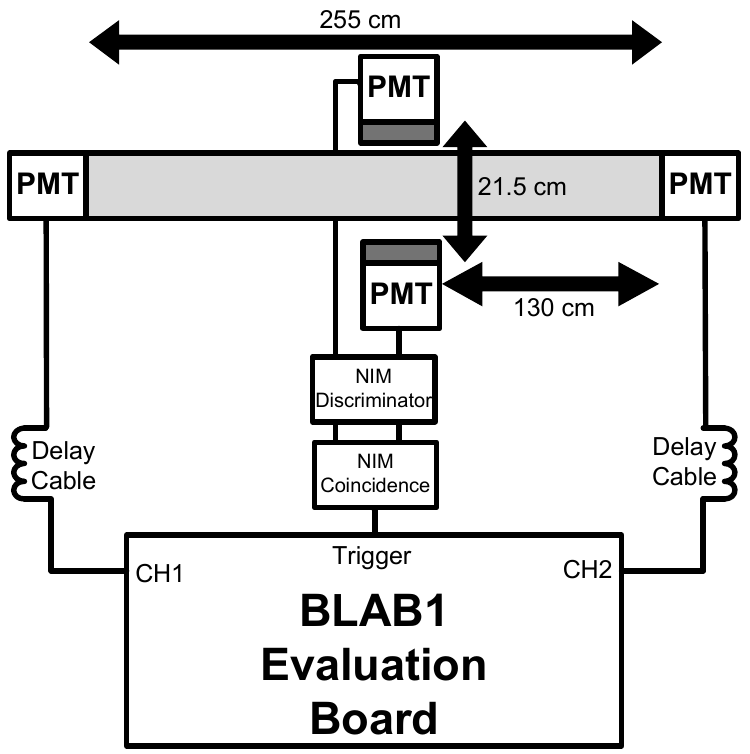}
\vspace*{0mm}
\caption{Schematic of the cosmic test system with a Belle TOF counter
and trigger counters.}
\label{belle_test}
\end{figure}

In order to evaluate the waveform sampler performance with a realistic
set of pulses, we use cosmic muons incident on a spare TOF counter of
the Belle detector~\cite{belle_tof}.  The test set-up is illustrated
in Fig.~\ref{belle_test}, and is located in the University of Hawaii
Instrumentation Development Laboratory.  

\begin{figure}[ht]
\vspace*{0mm}
\includegraphics[width=7.6cm]{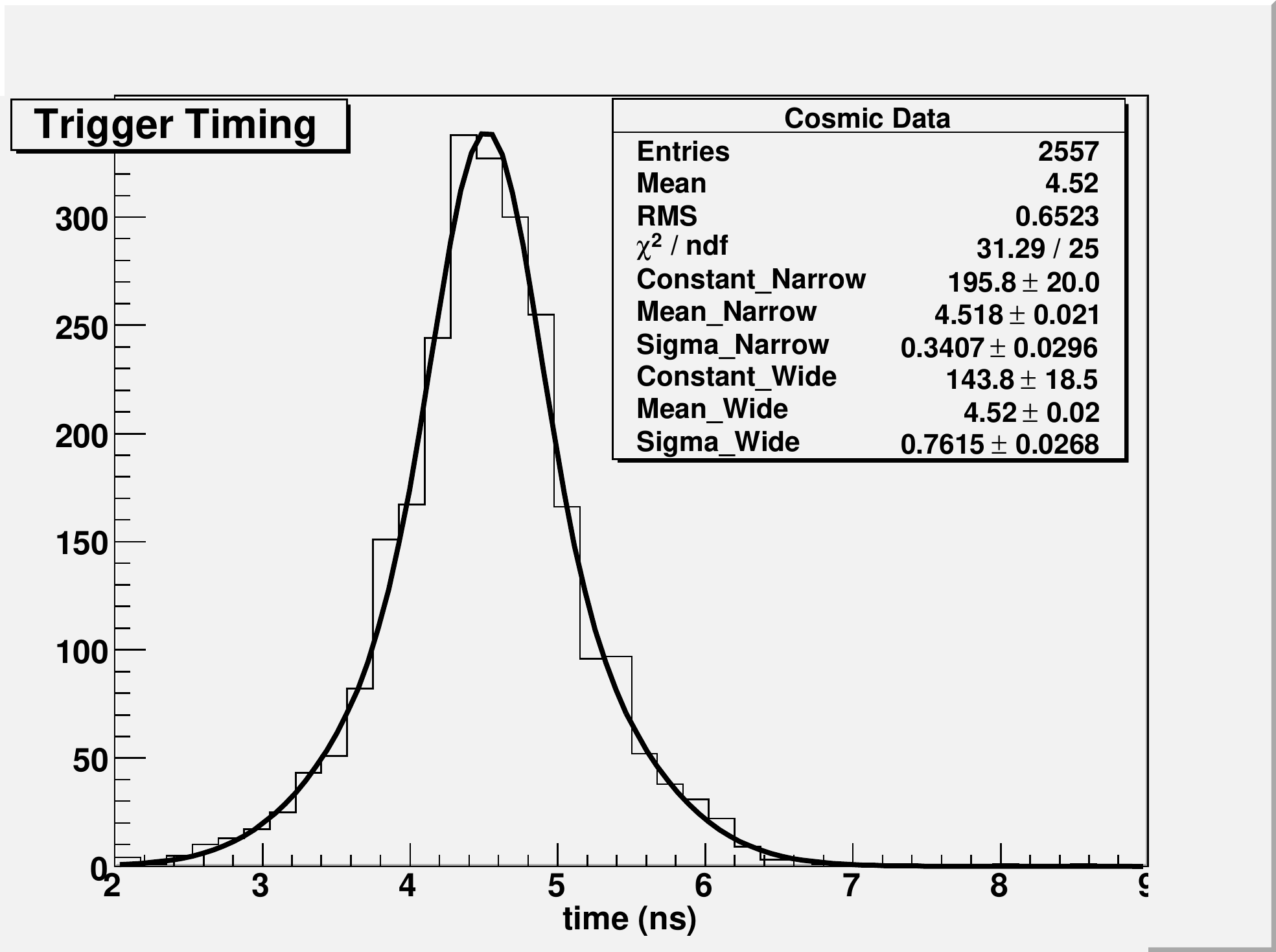}
\vspace*{0mm}
\caption{Timing results obtained for the trigger counter time difference.}
\label{trigger}
\end{figure}

A sample of a few thousand cosmic ray muons were recorded using the
test configuration shown in Fig.~\ref{belle_test}.  PMT signals from
both ends of the Bicron BC408 plastic scintillator bar are recorded.
The bar is 4cm thick, 255cm long and viewed by Hamamatsu R6680
fine-mesh PMTs at each end.  The Cherenkov trigger telescope counters
consists of lucite slabs (approx. 5cm x 6cm x 3.5cm), also viewed by
prototype R6680 fine-mesh PMTs.  To estimate expected system
performance, we recorded the trigger counters and extract an intrinsic
error on determination of the trigger time by comparing the observed
time difference in the two trigger counters.  This jitter, as shown in
Fig.~\ref{trigger}, is quite large and should be improved in the
future.

\begin{figure}[ht]
\vspace*{0mm}
\includegraphics[width=7.0cm]{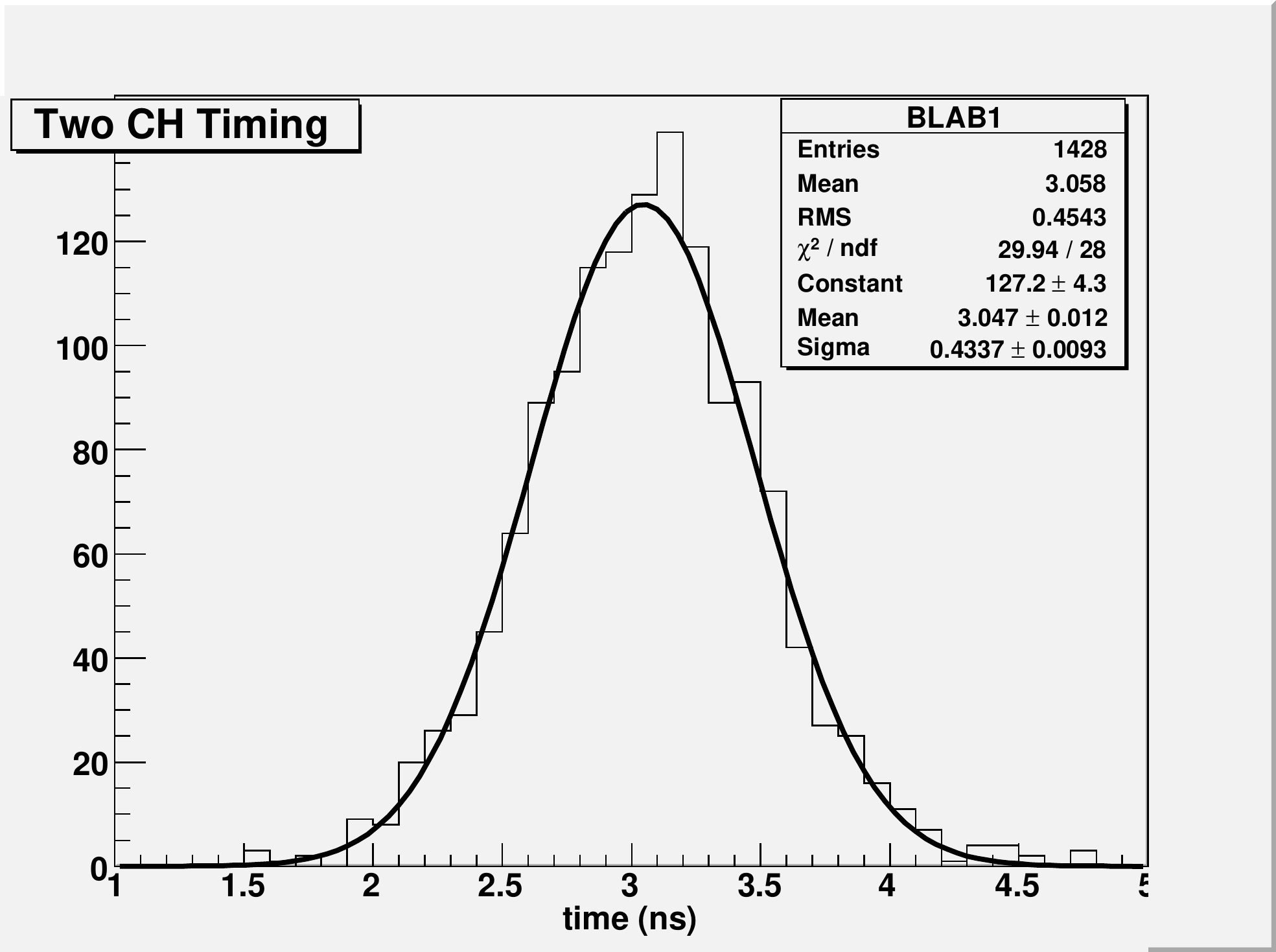}
\vspace*{0mm}
\caption{Timing difference results from fits to the PMT signal
waveforms at each end of the Belle TOF counter.}
\label{belle_results}
\end{figure}

The contribution of the narrow gaussian can be subtracted in
quadrature from the time difference observed at the ends of the Belle
TOF counter, the distribution of which is shown in
Fig.~\ref{belle_results}.

Doing this common mode subtraction leads to a resolution of about
190ps per PMT end.  Comparing this observed signal resolution with a
detailed Monte Carlo study \cite{jiwoo}, we can see that these values
are comparable to the 150ps (170ps) or so for single end times from MC
(data).



\section{Future Directions}
\begin{figure}[ht]
\vspace*{0mm}
\includegraphics[width=7.6cm]{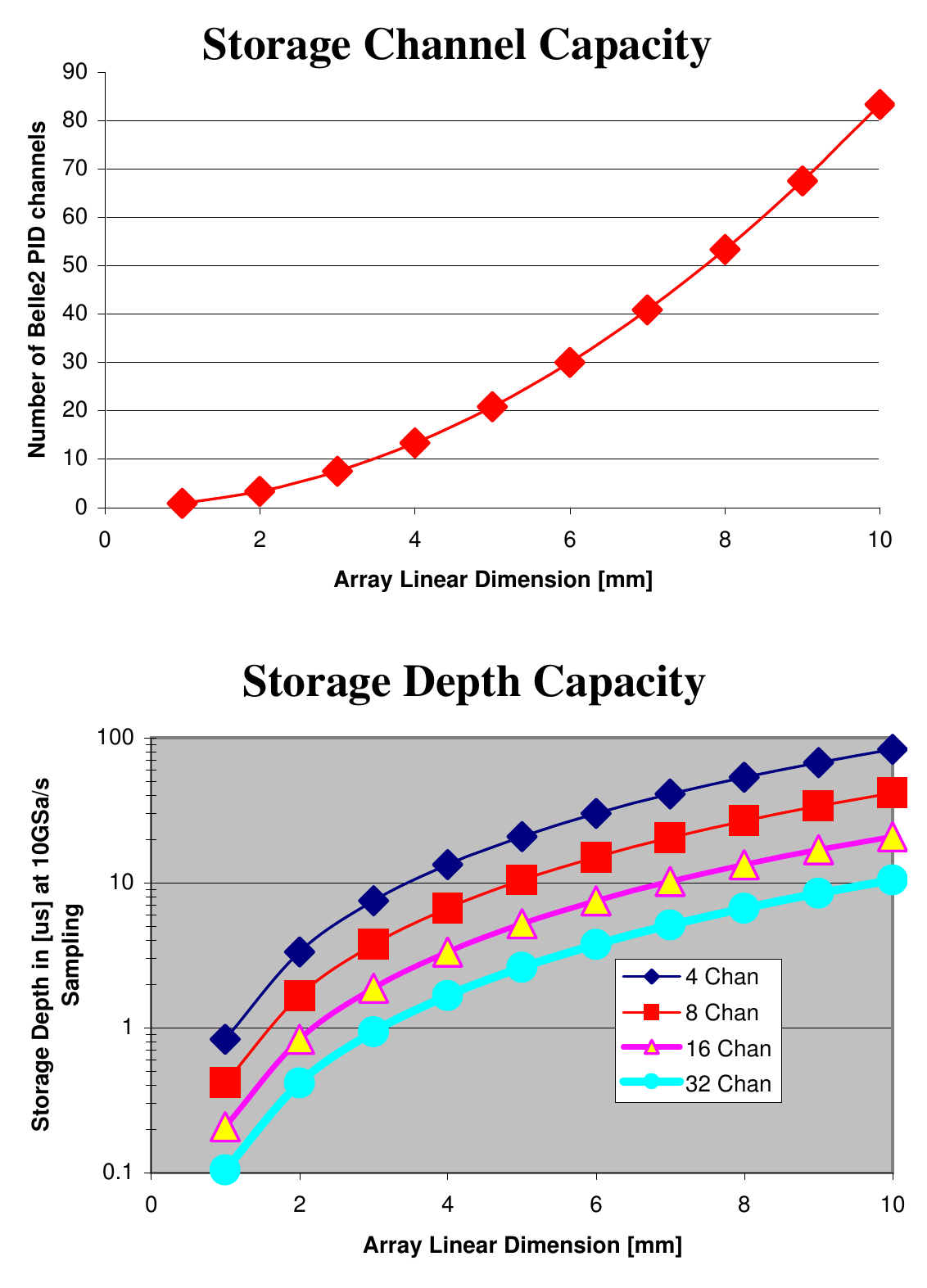}
\vspace*{0mm}
\caption{Packing density estimates for a future Belle upgrade Particle
Identification readout ASIC.  At top is number of channels versus
linear dimension of the (square) storage array, at bottom, the number
of $\mu $s of storage versus channel count also at 10GSa/s.  Pin
limitations will likely limit the practical number to 16 channels.}
\label{packing_density}
\end{figure}

While the analog bandwidth of the BLAB1 is adequate for many RF
recording applications, a higher bandwidth device will be explored,
based upon the lessons learned from this first device.  In particular,
the tree structure and design of the analog amplifier tree is being
scrutinized and improved in simulation.  It is hoped that an almost
arbitrarily large storage depth can be accommodated up to 1GHz of
analog bandwidth through a careful layout of the buffer amplifier
cascade array.  In future devices, it is possible to significantly
improve the number of storage cells.  A specific example of the
Particle Identification (PID) readout ASIC for the Belle upgrade is shown in
Fig.~\ref{packing_density}.

In the upper plot a $4\mu $s storage depth is assumed, or 40,000
storage cells at 10GSa/s.  In the lower plot, four separate curves
indicate the number of input channels and their subsequent depth
versus array linear distance (assumed to be square).  A die larger
than 1cm per side was not considered for yield reasons.  Also, pin
constraints, particularly on making the output parallel to reduce
readout latency, probably limit the practical number of input channels
to 16.  It is noted that many photodetectors operate at gains requiring
additional amplification in order to provide a signal with sufficient
amplitude for either triggering or recording.  Integration of
transimpedance and other input amplifier topologies are being studied
and results from future devices that use such on-chip, high analog
bandwidth elements will be reported later.

\begin{figure}[ht]
\vspace*{0mm}
\includegraphics[width=7.6cm]{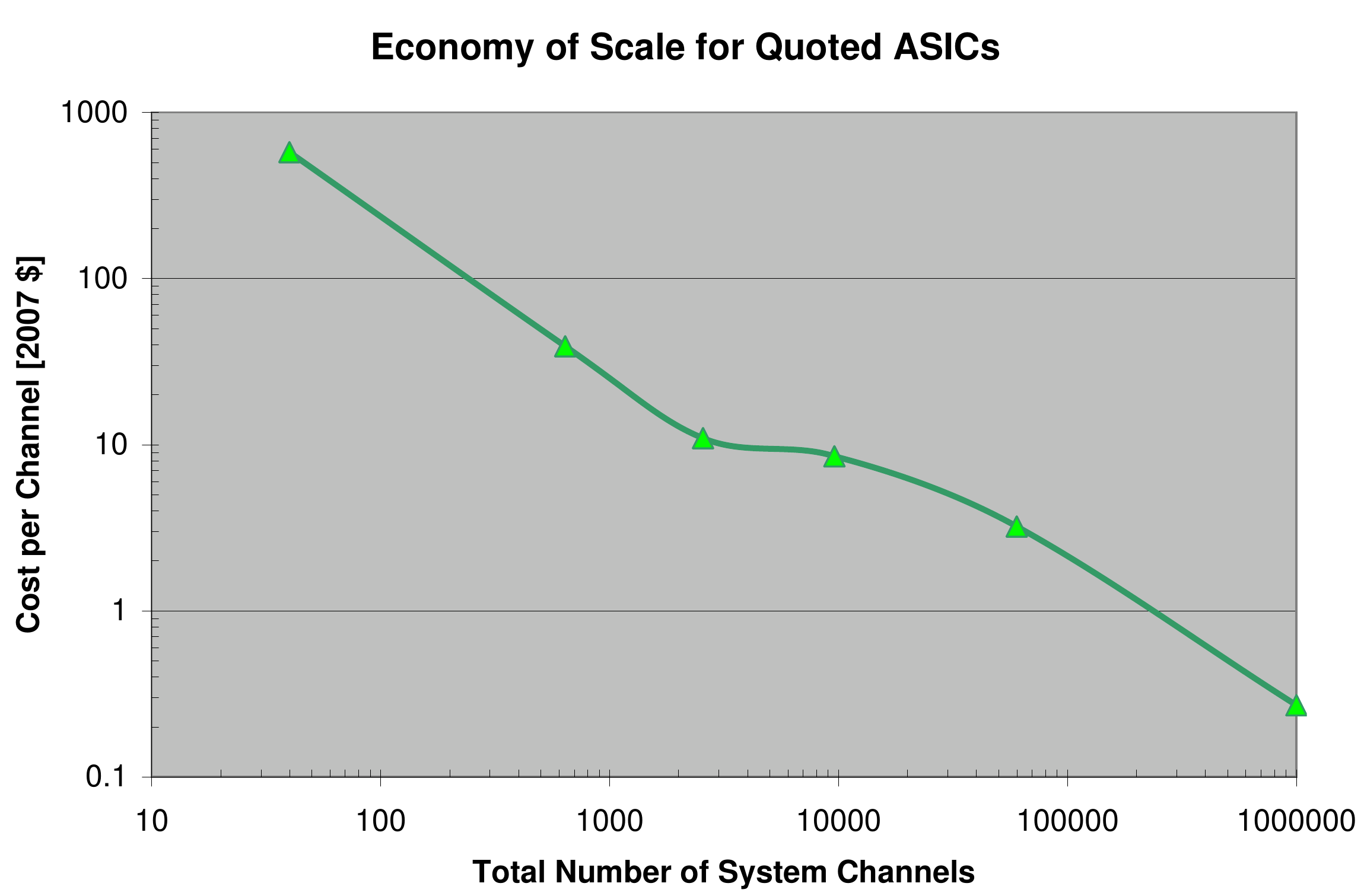}
\vspace*{0mm}
\caption{Channel cost scaling for a reference waveform sampler ASICs
based upon recent experience.}
\label{economy_scale}
\end{figure}

There is a misconception that waveform sampling is significantly more
expensive than traditional discriminator + TDC methods.  Certainly
when packaged as a full oscilloscope and sold as a commercial unit,
with large buffer depth, this can be true.  In
Fig.~\ref{economy_scale} is listed the fabricated and quoted prices
during the 2007 fiscal year in the same TSMC 0.25$\mu $m process.  It
is interesting to note that the slope of the first 3 devices
correspond to Multi-Project Wafer runs, and the latter 3 are dedicated
wafer runs.  Packaging is not included and is a minimum of about
\$1/die in high volume.

A summary of active ASIC designs inspired by the performance of the
BLAB1 ASIC may be found in Table~\ref{future}.


\begin{table}[hbt]
\caption{\it Future BLAB1 inspired ASIC designs.}
\label{future}
 \begin{center}
    \begin{tabular}{|l|c|c|c|c|c|} \hline
{\it ASIC }& {\it \# } & {\it Samples} & Rate  & {\it BW } & {\it
power } \\ 
{\it Acronym }& {\it Chan } & {\it per Chan } & {\it [GSa/s] } & {\it [GHz] } & {\it
mW/chan } \\ \hline\hline 
BLAB2 & 16 & 2k & 2-10 & $\geq $ 1 & $\leq 20$ \\ \hline 
TARGET & 16 & 4k & 0.5-1 & 0.5 & $\leq 20$ \\ \hline 
RAL64 & 64 & 512 & DC - 5 & $\approx $ 0.3 & $\leq 10$ \\ \hline 
APTD & 4 & 8k & DC - 0.5 & 0.2 & $\leq 1$ \\ \hline 
BIRD & 1 & 256k & 1 & 0.5 & $\leq 30$ \\ \hline \hline
     \end{tabular}
  \end{center}
\end{table}   
These devices find application niches for the following reasons:

\vspace{0.1in}

\begin{itemize}
\item {\bf Timing Performance $\rightarrow $ } {\tt BLAB2} is intended for sub-10ps
photodetector pulse time recording
\vspace{0.05in}
\item {\bf Low Cost $\rightarrow $ } {\tt TARGET} is intended for the
low-cost instrumentation of 1M photodetector channels of a future TeV
$\gamma $ telescope
\vspace{0.05in}
\item {\bf High Density $\rightarrow $ } {\tt RAL64} is a dense array readout device, where
128 channels or more could be considered in the future
\vspace{0.05in}
\item {\bf Low Power $\rightarrow $ } {\tt APTD} is a demonstrator low-power ADC device for a
proposed Advanced Pair Telescope satellite
\vspace{0.05in}
\item {\bf Extended Depth $\rightarrow $ } {\tt BIRD} is a very deep storage ASIC for the
future IceRay extended radio neutrino detector at the South Pole 
\end{itemize}

\vspace{0.1in}

A number of these designs are reaching maturity and two have already
been submitted for fabrication.  Details of the designs and results
from operation of these devices will be reported in the future.


\section{Summary}
A first generation of deep-storage Switched Capacitor Array (SCA) CMOS
device has been studied in a 0.25$\mu$m process.  This architecture is
optimized for concurrent acquisition and readout, permitting
deadtimeless operation.  Demonstrated low-power, high-resolution and
exquisite timing performance make this device and subsequent variants
attractive for readout of a broad range of particle and astroparticle
detectors.


\section{Acknowledgements}
The authors gratefully acknowledge the generous support of the MOSIS
Educational Program, which provided the fabrication of the BLAB1 ASIC
prototype through their University Research Program.  Testing was
supported in part by Department of Energy Advanced Detector Research
Award \# DE-FG02-06ER41424.



\begin{thebibliography}{99}
\bibitem{national_semi} For example the ADC08D1500 from National
Semiconductor, which digitizes two channels at 1.5GSa/s to 7.2
effective bits of resolution while consuming 1.9Watts.
\end{thebibliography}
\end{document}